\pdfoutput=1
\documentclass[epjST-vaso,a4paper,twoside,instindent,runningheads]{svjour-vaso} 

\usepackage{graphicx}        

\usepackage{amsmath,amssymb}
\usepackage{url,nicefrac,siunitx,xspace,textcomp,cite,color,xcolor,orcidlink}
\definecolor{darkgreen}{cmyk}{0.9, 0, 0.9, 0}
\definecolor{darkred}{cmyk}{0, 0.6, 0.6, 0}
\usepackage[labelformat=simple,list=true]{subcaption}

\usepackage{hyperref}
\hypersetup{
  colorlinks=true,
  linkcolor=blue,   
  citecolor=darkgreen,   
  urlcolor=blue,     
  citebordercolor=0 1 0,
  filebordercolor=0 1 1,
  linkbordercolor=1 0 0,
  menubordercolor=1 0 0,
  urlbordercolor=1 0 0,
  runbordercolor=0 .7 .7,
}

\DeclareGraphicsExtensions{.pdf,.png,.jpg,.eps}
\graphicspath{{./}{./figs/}}

\newcommand{\be}{\begin{equation}}
\newcommand{\ee}{\end{equation}}
\newcommand{\ifb}{\ensuremath{\mathrm{fb}^{-1}}\xspace}
\newcommand{\ipb}{\ensuremath{\mathrm{pb}^{-1}}\xspace}
\newcommand{\tev}{\mbox{TeV}\xspace}
\newcommand{\gev}{\mbox{GeV}\xspace}
\newcommand{\cmssr}{\ensuremath{\mathrm{cm^{-2} s^{-1} sr^{-1}}}\xspace}
\newcommand{\gd}{\ensuremath{g_{\mathrm D}}\xspace}
\newcommand{\dedx}{\ensuremath{{\mathrm d}E/{\mathrm d}x}\xspace}
\newcommand{\hf}{\nicefrac{1}{2}\xspace}
\newcommand{\bt}{\ensuremath{\hat{\beta}}\xspace}
\newcommand{\nova}{NO\ensuremath{\nu}A\xspace}

\begin{document}

\title{Magnetic monopoles: from Dirac to the Large Hadron Collider}

\author{Vasiliki A.\ Mitsou\,\orcidlink{0000-0002-1533-8886}\thanks{vasiliki.mitsou@ific.uv.es}} 

%
%



\institute{Instituto de F\'isica Corpuscular (IFIC), CSIC -- Universitat de Val\`encia, \\ C/ Catedr\'atico Jos\'e Beltr\'an 2, 46980 Paterna (Valencia), Spain}
%

\abstract{One of the basic properties of magnetism is that a magnet has always two poles, north and south, which cannot be separated into isolated poles, the magnetic monopoles. There are strong theoretical arguments in favour of monopoles' existence, but in spite of extensive searches they are yet to be found. In this review article, after highlighting briefly the theoretical foundations of monopoles, a historical overview of experimental endeavours to observe them is given, with emphasis on the state-of-the-art of searches in cosmic and collider experiments and in particular the Large Hadron Collider at CERN. 
}


\maketitle

\newpage

\setcounter{secnumdepth}{5}
\setcounter{tocdepth}{4}
\tableofcontents

\newpage
\section{Historical overview}\label{sec:intro}

The principal theoretical motivation behind the introduction of magnetic monopoles (MMs) is the symmetrisation of the Maxwell's equations and the explanation of the charge quantisation. Maxwell’s original formulation of Electromagnetism consisted of 20~variables in 20~equations. Oliver Heaviside in 1884 introduced the \emph{curl} and \emph{divergence} operators from vector calculus to rewrite these equations into the form by which they have been known ever since. When Heaviside produced his variant of the Maxwell’s equations, he included magnetic charges called \emph{magnetons}.
When he finished he just set the magnetic charge to zero by hand to account for the non-observation of magnetic charges. In 1894, Pierre Curie was the first to suggest that \emph{free magnetic monopoles} could exist~\cite{Curie:1894}.

In 1931 Paul Dirac showed that the existence of MMs in the Universe could offer an explanation for the discrete nature of the electric charge, leading to the Dirac Quantisation Condition (DQC)~\cite{Dirac:1931kp,Dirac:1948um}:
\be 
e g = \frac{n}{2}, \quad n \in \mathbb{Z}, 
\label{eq:dqc}
\ee
\noindent where $e$ is the electron charge and $g$ is the MM magnetic charge. Natural units, namely, $\hbar = c = 4\pi\varepsilon_0 = 1$, where $\hbar$ is the reduced Planck constant, $c$ the speed of light in the vacuum and $\varepsilon_0$  the vacuum permittivity, are used throughout the review. The electron charge is related to the fine structure constant through the expression $\alpha = e^2 \simeq \frac{1}{137}$.\footnote{The fine structure constant at zero energy is considered in general, as appropriate to the fact that the DQC pertains to long (infrared) distances from the centre of the monopole. However, in cases of production at particle collisions, the $\alpha$ value at the centre-of-mass energy may be more suitable.}  
This quantisation condition should be modified by a factor of three if free quarks are found. The MM mass and spin remain free parameters of the theory.

This attractive proposal provoked a number of experimental and observational investigations since then, which are reviewed here. These lecture notes are organised as follows. Section~\ref{sc:theo}  serves as a brief reference of the various theoretical scenarios predicting the existence of MMs. The MM interactions with matter and associated detection techniques are discussed in Section~\ref{sc:techniques}. The state-of-the-art in monopole searches is discussed in Section~\ref{sc:cosmics} for MMs of cosmic origin and in Section~\ref{sc:colliders} for experiments in colliders. Many of the searches for MMs are also sensitive to particles with high electric charges; resulting constraints are presented in Section~\ref{sc:hecos}. Section~\ref{sc:cosmic-colliders} discusses the relevance of collider searches for MMs coming from the sky, along with ensuing results. The review concludes with insights on future directions in Section~\ref{sc:future} and an epilogue in Section~\ref{sc:summary}.

\section{Theory}\label{sc:theo}

Magnetic monopoles~\cite{Preskill:1984gd,Shnir:2005xx,Rajantie:2012xh,Mavromatos:2020gwk}, carrying isolated magnetic charges, analogous to electrically charged particles, have been hypothesised over the years in numerous theoretical proposals. Some of the theories predicting the existence of a MM and objects related to MMs are highlighted below, while a comprehensive review is given in Ref.~\cite{Mavromatos:2020gwk}.

\subsection{Dirac's monopole}\label{sc:dirac}

In Dirac's formulation~\cite{Dirac:1931kp,Dirac:1948um}, MMs are assumed to exist as point-like particles and quantum-mechanical consistency conditions lead to \eqref{eq:dqc}, establishing the value of their magnetic charge. Although MMs symmetrise Maxwell's equations in form, there is a numerical asymmetry arising from the DQC, namely, that the basic magnetic charge is \emph{much larger} than the smallest electric charge. A MM with a single \emph{Dirac charge} $\gd\equiv 1/(2e) = e/(2\alpha)$ has an equivalent electric charge of  $\sim137 e/2=68.5e$. 

\subsection{GUT monopoles}\label{sc:gut}

Efforts to search for MMs have been motivated considerably, since Grand Unified Theories (GUTs) of strong and electroweak interactions predicted the existence of MMs. In 1974, 't Hooft~\cite{tHooft:1974kcl} and Polyakov~\cite{Polyakov:1974ek}, independently of each other, pointed out that a unified gauge theory in which Electromagnetism is embedded in a semi-simple gauge group, such as $SU(2)$, would require the existence of the MM as a soliton with spontaneous symmetry breaking. To be more specific, a semi-simple non-abelian gauge group may break into its subgroups including $U(1)$ which essentially describes MMs in the framework of GUTs. GUT monopoles are far too massive be produced at any foreseeable human-made accelerator, having a predicted mass of ${\mathcal O}(10^{15}~\gev)$ or higher, leaving the cosmically created MMs as  the only means to observe them.

\subsection{Electroweak monopole}\label{sc:cho}

The \emph{electroweak} monopole, proposed by Cho and Maison~\cite{Cho:1996qd,Cho:2012bq,Cho:2019vzo}, is a generalisation of the Dirac MM, representing a non-trivial hybrid of Dirac and 't Hooft--Polyakov MM that carries magnetic charge twice that of the Dirac monopole, $2\gd$. This is because it is based on the quotient group $SU(2) \times U(1)_Y / U(1)_\text{em} $, where $U(1)_\text{em}$ is the (unbroken) group of Electromagnetism instead of, e.g., the $SU(2)$ group in the Georgi--Glashow model. Born--Infeld regularisation has been used to create finite-energy, stable solutions~\cite{Arunasalam:2017eyu,Ellis:2016glu,Mavromatos:2018kcd,Farakos:2025byy,Mavromatos:2026nlp} for the ---initially characterised  by core singularities--- electroweak monopole. Estimates of the electroweak MM mass indicated that it may be as low as $\sim5.5~\tev$ or less, thus allowing its potential pair production at the LHC~\cite{Ellis:2016glu}. 

Earlier, Lee and Weinberg~\cite{Lee:1994sk} had shown that a finite-energy MM can exist without requiring a non-trivial topology. The crucial point is that the charged vector boson field $W_\mu$ develops a magnetic moment, which can screen the singular magnetic field near the MM core. Thus, the finite energy arises from fieldd dynamics rather than topology.   

P.~Q.~Hung proposed a model where topologically stable, finite-energy MMs \`a la 't Hooft--Polyakov could exist with a mass proportional to the electroweak scale. This comes about in a model of neutrino masses where right-handed neutrinos are \emph{non-sterile}, whose electroweak-scale Majorana masses are obtained by the coupling to a complex triplet Higgs field~\cite{Hung:2020vuo,Ellis:2020bpy}. Unlike scenarios with sterile neutrinos, this class of models can be probed at particle colliders through unconventional signatures involving long-lived particles~\cite{Mitsou:2026ney}.

\subsection{Dyons}\label{sc:dyon}

Dyons, possessing both magnetic and electric charge, were proposed in 1969 by Schwinger~\cite{Schwinger:1969ib} as a more complex solution to the DQC, which depends on the underlying theoretical scenario. The DQC for two dyons of electric charge $e_1, e_2$ and magnetic charge $g_1, g_2$ reads:
\be\label{eq:dqc-dyon}
e_1 g_2 -  e_2 g_1 = \frac{n}{2}, \quad n \in \mathbb{Z}.
\ee
Most arguments made for MMs may be naturally extended to dyons. The existence of both monopoles and dyons is predicted in GUTs~\cite{Witten:1979ey,Dokos:1979vu,LYi:1982ibm}, in Einstein--Yang--Mills theories~\cite{Bjoraker:1999yd}, Kaluza--Klein theories~\cite{Sen:1997zb}, and string theory~\cite{Dabholkar:2008zy}. 

Dyons may play a role in the way Charge-conjugation--Parity (CP) symmetry is violated in gauge theories. It has been demonstrated through semi-classical arguments that in CP-conserving theories the dyon charge is quantised~\cite{Tomboulis:1975qt,Hasenfratz:1976vb,Christ:1976cg}. Through the Witten effect~\cite{Witten:1979ey}, dyons directly link the CP-violating vacuum angle of a theory to the electric charge. Since it has been found experimentally that CP is violated, this effect predicts that the dyon electric charge should deviate from  an integer multiple of the electron's charge by a small amount.

\subsection{Global monopole}\label{sc:global}

\emph{Global} monopoles have been proposed~\cite{Barriola:1989hx} as spacetime (cosmological) defects allowing for the spontaneous breaking of internal global $SO(3)$ symmetries in non-gauged Georgi--Glashow models. These MMs carry \emph{no magnetic charge}; however, their gravitational effects far away from their centre are significant, in the sense that the (non-Minkowski) spacetime has a deficit angle. Although minute, such an effect affects the forward scattering amplitude of Standard Model (SM) particles that propagate in such backgrounds, leading to ring-like angular regions, where the scattering amplitude is very large~\cite{Mazur:1990ak,Mavromatos:2017qeb}. Such \emph{peculiar scattering patterns} of ordinary SM particles may indicate indirectly the presence of a neutral global monopole in collider detectors, where they may pair-produced~\cite{Barriola:1989hx,Drukier:1981fq}. Furthermore, a variant of the global monopole model, including axion fields (with a given \emph{axion charge}) and a real electromagnetic field, which couples only gravitationally to the scalar $SO(3)$ symmetry breaking sector, has been considered~\cite{Mavromatos:2016mnj,Mavromatos:2018drr,Chatzifotis:2022ubq,Mavromatos:2026acc}, resulting into axions capable of inducing electromagnetic MM solutions with a \emph{real magnetic charge} of order of the axion charge. In such a case, \emph{both} the high ionisation \emph{and} the peculiar effects~\cite{Mazur:1990ak,Mavromatos:2017qeb} of the monopole background on the scattering of ordinary particles on them are in operation. 

\subsection{Monopolium}\label{sc:monopolium}

A possible explanation for the lack of experimental confirmation of MMs is Dirac's~\cite{Dirac:1931kp} and Zeldovich's~\cite{Zeldovich:1978wj} proposals that MMs are not seen freely, because they form a bound state called \emph{monopolium}~\cite{Nambu:1977ag,Hill:1982iq,Dubrovich:2003mv,Vento:2007vy,Epele:2007ic,Epele:2008un,Abreu:2020qqy} being confined by strong magnetic forces. Most of the monopolium studies follow the low-energy effective theory of Ginzburg and Schiller~\cite{Ginzburg:1998vb,Ginzburg:1999ej}. This is based on the standard electroweak theory, where the MM is coupled to the photon and weak bosons assuming that its mass is much larger than the $Z$ mass and that the MM interacts with the fundamental fields of the $SU(2) \times U(1)$ theory before symmetry breaking.  More information on the experimental signatures and constraints are given in Section~\ref{sc:box}.

\section{Detection techniques}\label{sc:techniques}

The detection of MMs relies on various types of interactions with matter, which are briefly discussed in this section. It is stressed that the same technique may be applied to MMs of different origins, i.e.\ for cosmic rays or when produced in colliders. However, the sensitivity ---thus the detector optimisation--- may vary and depend, e.g.\ on the MM velocity and mass.  To these techniques, one can add the application of the Gauss's law for magnetism to sense the presence of net magnetic charge in a (large) volume. 

\subsection{Ionisation and excitation}\label{sc:ionis}

Particles carrying electric or magnetic charge will deposit some amount of energy through the ionisation process and excitation of atoms when they traverse matter. To calculate the energy loss of MMs passing through matter, first the energy-transfer process of the MM to the surrounding medium has to be specified. There are three primary ways via which the energy can be dissipated and their importance depends on the MM velocity $\beta$  and the medium~\cite{Giacomelli:2003yu}.

\begin{description}
\item[\textbf{Ionisation}] The energy transferred leads to the production of free electrons. This is given by the Bethe--Bloch formula discussed below and is the dominant energy-loss mechanism for fast monopoles $(\beta>0.05)$ moving in gaseous detectors.
\item[\textbf{Atomic excitation}] The energy is transferred to atoms of higher energy states. It starts to dominate the energy loss at slower MMs, i.e.\ for $10^{-3}\lesssim\beta\lesssim10^{-2}$.
\item[\textbf{Elastic collisions with atoms}] The energy loss is due to atoms (nuclei) recoiling through the MM coupling to atomic or nuclear magnetic moment. It becomes important for even slower MMs  $(\beta\lesssim10^{-3})$ and different energy-loss calculations have been performed for diamagnetic and paramagnetic materials.
\end{description}
 
The energy loss per path length travelled due to ionisation, \dedx, for a particle with electric charge $ze$ is well-described by the Bethe--Bloch formula~\cite{Bethe:1930ku,ParticleDataGroup:2024cfk}, which for high energies takes the form:
\be\label{eq:bethe-bloch}
-\frac{{\mathrm d}E}{{\mathrm d}x} =  K\frac{Z}{A} \frac{z^2}{\beta^2} \left[ \frac{1}{2} \ln \left( \frac{2m_e\beta^2\gamma^2 T_\text{max}}{I^2} \right)
 - \beta^2 - \frac{\delta(\beta\gamma)}{2} \right], 
\ee
\noindent where $K=4\pi N_{\mathrm A} r_e^2 m_e$, with $m_e$ $(r_e)$ the electron mass (radius), $N_{\mathrm A}$ the Avogadro number;  $\gamma=1/\sqrt{1-\beta^2}$ is the Lorentz factor; and $Z$, $A$, and $I$ are the atomic number, mass number and mean excitation energy of the medium, respectively. The maximum kinetic energy transferred to a free electron in a single collision is expressed by $T_\text{max}$, and $\delta(\beta\gamma)$ is a density effect correction.

The stopping power formula~\eqref{eq:bethe-bloch} has been adapted by Ahlen~\cite{Ahlen:1978jy,Derkaoui:1998uv} to describe the ionisation energy loss for a magnetic charge $g=n\gd$ to the expression:
\be\label{eq:ahlen}
-\frac{\textrm{d}E}{\textrm{d}x}=K\frac{Z}{A}g^2\left[\ln\frac{2m_e\beta^2\gamma^2}{I}+\frac{K(g)}{2}-\frac{1}{2}-B(g)-\frac{\delta(\beta\gamma)}{2}\right] ,
\ee
where $K(g)$ is the Kazama--Yang--Goldhaber correction~\cite{Kazama:1976fm} which is the result of the relativistic cross-section calculations and $B(g)$ is the Bloch correction.

\begin{figure}[htb]
\centering
     \begin{subfigure}[b]{0.47\textwidth}
         \centering
	\includegraphics[width=\textwidth]{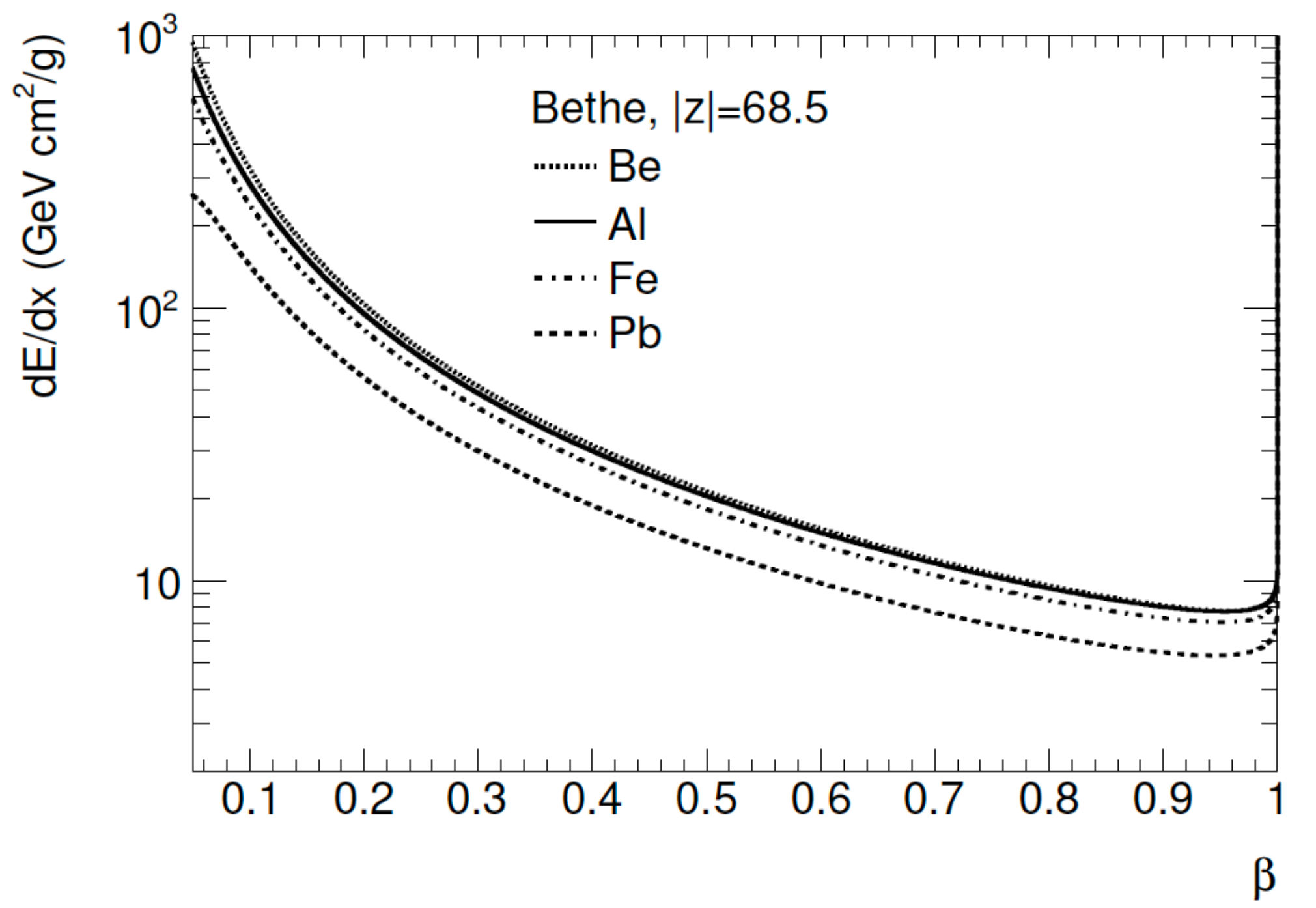} 
	\caption{\dedx for electric charges.}\label{fig:dEdx_electric}
     \end{subfigure}	
\hspace{0.01\textwidth}
     \begin{subfigure}[b]{0.47\textwidth}
         \centering
	\includegraphics[width=\textwidth]{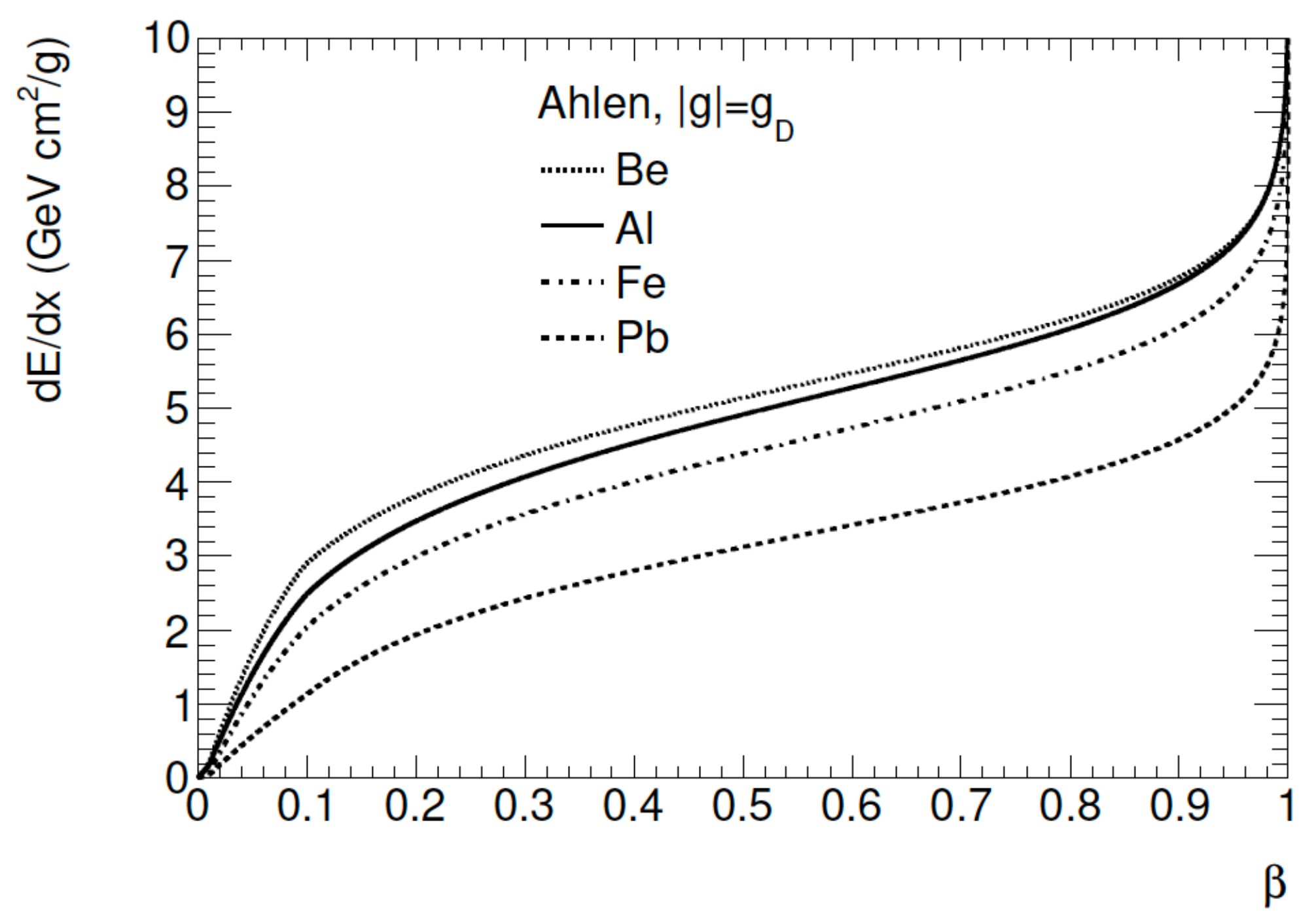} 
    	\caption{\dedx for magnetic charges.}\label{fig:dEdx_magnetic}
     \end{subfigure}		
 \caption{Ionisation energy loss \dedx in various materials as a function of velocity $\beta$ for HIPs possessing \protect\subref{fig:dEdx_electric} an electric charge of $68.5e$, and \protect\subref{fig:dEdx_magnetic} a magnetic charge of 1\gd, based on~\ref{eq:bethe-bloch} and~\ref{eq:ahlen}, respectively. For  low-$\beta$ magnetic charges in \protect\subref{fig:dEdx_magnetic}, estimates from Ref.~\cite{Ahlen:1982mx} are used.  From Ref.~\cite{DeRoeck:2011aa}. }\label{fig:dedx}
\end{figure}

For instance, a relativistic MM with charge \gd\ loses energy as a nucleus with $z \simeq 69$, or $(\gd/e)^2 = 68.5^2 \simeq 4,\!700$ times more than an electron. This makes even a singly charged MM a highly ionising particle (HIP). Equation~\ref{eq:ahlen} is only valid for $\beta \gtrsim 0.1$; for slower MMs with $\beta \lesssim 0.01$, \dedx falls linearly as the MM velocity decreases~\cite{Ahlen:1982mx,Derkaoui:1998uv}. The different dependence of energy loss on the MM velocity between electric and magnetic charges is depicted in Figs.~\ref{fig:dEdx_electric} and~\ref{fig:dEdx_magnetic}, respectively, for the same equivalent charge. Evidently, there is no Bragg peak for MMs and their range in the absorbing volume is consistently longer than for electric charges.

Through the ionisation process and excitation of atoms, liquid or plastic scintillators, gas detectors, and nuclear track detectors are sensitive to MMs, as well as to slow-moving objects with multiple electric charges. One disadvantage of these detection techniques is that it is in general difficult ---yet not impossible (cf.\ Section~\ref{sc:ntd})--- to disentangle magnetic from high electric charges solely from the energy deposit information. Therefore, a careful review of the radiation environment of the setup is necessary in order to exclude the possibility that the candidate event is due to background sources, e.g.\ from heavy-ion fragments. On the other hand, hypothetical particles characterised by \emph{electric charge} can also be probed with such experimental setups, as we shall see in Section~\ref{sc:hecos}.

\subsubsection{Scintillators}\label{sc:scint}

Atom excitation induced by slow MMs passing through a scintillation counter invokes a light yield much larger than that of a minimum ionising particle. The light yield is saturated for velocities  $10^{-3}\lesssim\beta\lesssim10^{-1}$, while it increases again for  $\beta > 0.1$ due to secondary emission of $\delta$-rays~\cite{Patrizii:2015uea}.

An extension of MM searches to even lower velocities would require detection materials with small band gaps, e.g.\ acrylic-based scintillators with large concentrations of naphthalene are probably sensitive to MMs down to $5 \times 10^{-4}$. With a band gap of only $\sim1.1$~eV, silicon detectors would be able to detect MMs moving as slowly as $10^{-4}$~\cite{Ahlen:1983rw}.

\subsubsection{Gaseous detectors}\label{sc:gas}

For velocities $\beta\gtrsim 10^{-3}$ the high ionisation in drift and streamer tubes provides a good handle on discerning MMs from minimum ionising particles, such as muons. For slower MMs and for gases such as hydrogen and helium, the Drell~\cite{Drell:1982zy} and Penning~\cite{Penning:1927} effects can be exploited: a MM leaves the atoms in a metastable excited state and the MM can be detected by the radiation produced by the subsequent return of electrons to their ground state.

The Drell mechanism is effective as long as the MM--atom collision energy exceeds the atomic-level energy split. Detection via this  effect may be based either on the observation of the photon emitted  during the de-excitation of the excited atom or by observing the ionisation caused by the energy transfer from the excited atoms to complex molecules with a small ionisation potential
(Penning effect)~\cite{Giacomelli:1984gq}.

\subsubsection{Nuclear track detectors}\label{sc:ntd}

\begin{figure}[htb]
\centering
\begin{minipage}[b]{0.47\linewidth}
\centering
  \includegraphics[width=0.78\textwidth]{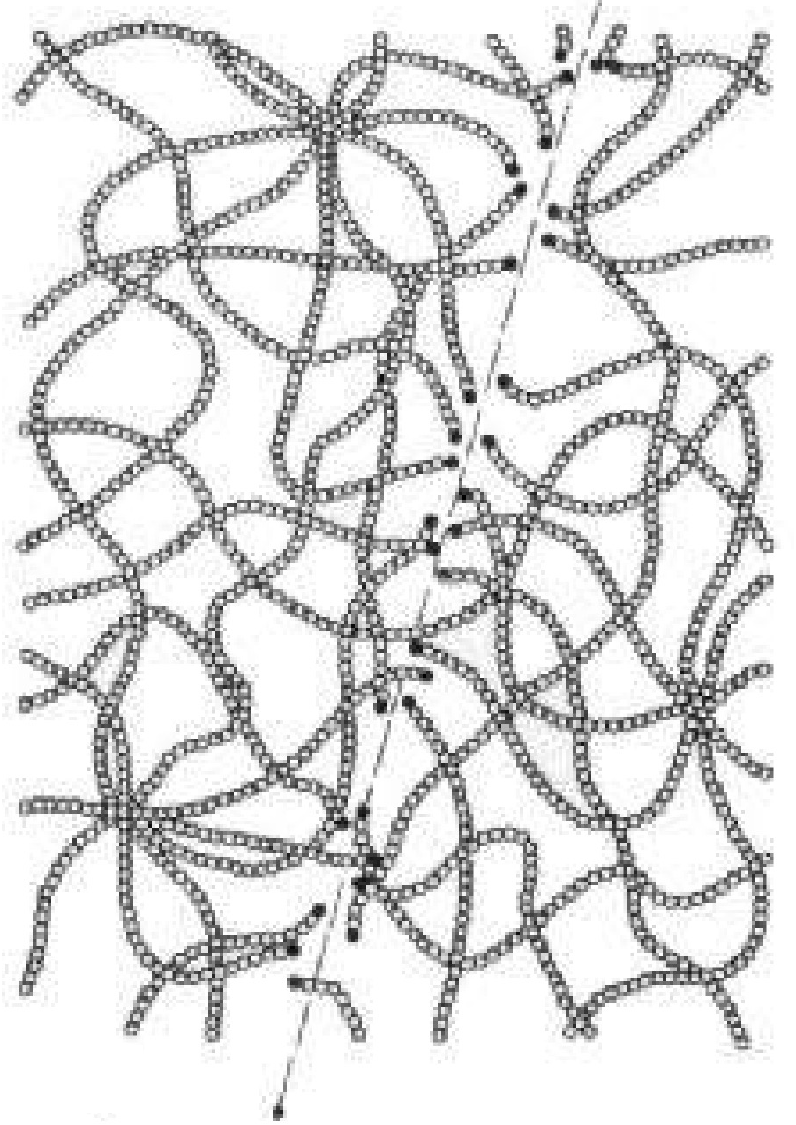}
  \caption{Breaking of the polymeric bonds by a crossing charged particle. The damaged region forms a latent track around the particle trajectory. From Ref.~\cite{Fleischer:1969en}.} \label{fig:polymeric}
\end{minipage}   
\hspace{0.04\linewidth}
\begin{minipage}[b]{0.47\linewidth}
\centering
  \includegraphics[width=0.63\textwidth]{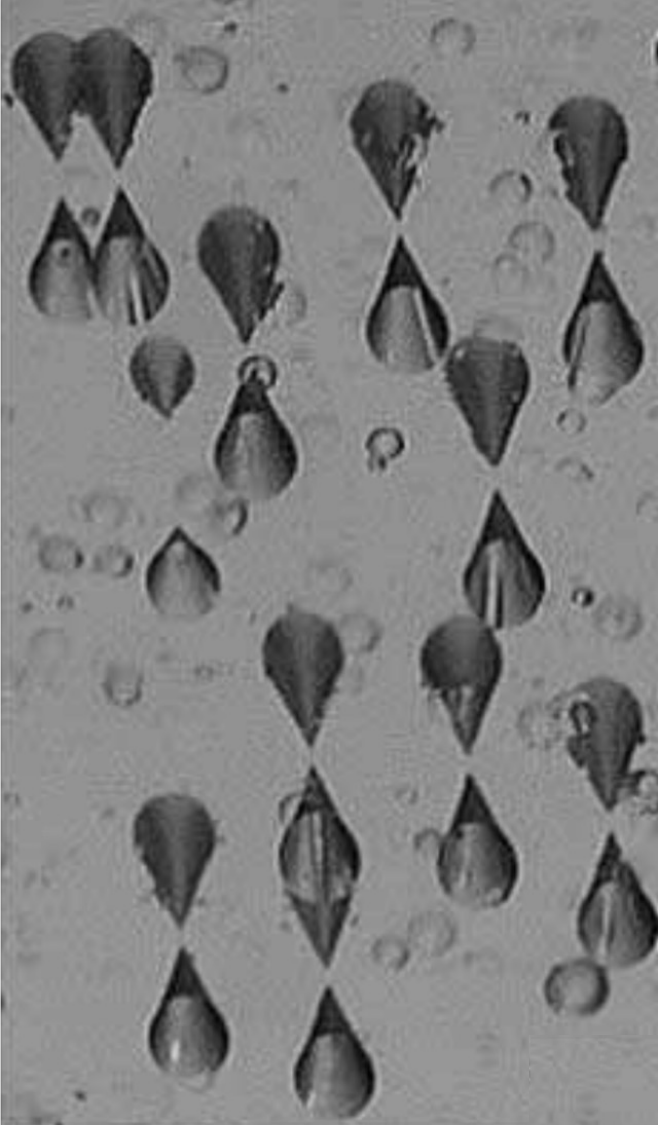}
  \caption{Etch cones on a CR-39\textsuperscript{\textregistered} foil formed from 400~$A$ MeV Fe$^{26+}$ ions and their fragments after strong etching: 8N KOH + 1.25\% Ethyl alcohol at 77\unit{\degreeCelsius} for 30~h. From Ref.~\cite{Giorgini:2007fm}.} \label{fig:etchpit}
\end{minipage}
\end{figure}

When traversing nuclear track detector (NTD) panels, HIPs such as MMs or electric charges damage the material  at the level of polymeric bonds within a cylindrical region extending to a few tens of nanometers around the particle trajectory, forming a \emph{latent track}~\cite{NIKEZIC200451}, as shown in Fig.~\ref{fig:polymeric}. The latter is related to the \emph{restricted energy loss}, which is the fraction of the total energy loss localised in this cylindrical region. When the NTD sheets are chemically etched after exposure, the latent tracks are revealed as cone etch pits ---examples are shown in Fig.~\ref{fig:etchpit}--- which can then be identified after proper scanning of the NTDs. 

The identification of etch-pit candidates during scanning is an involved procedure and improvements are being proposed to facilitate it. NTD sheets may become opaque after chemical etching due to  high-radiation environment, thus complicating the track identification. By deploying liquid thin films on both sides of the irradiated foils and using machine-vision techniques, etch-pits  can be detected and classified~\cite{Kalliokoski:2026yjm}. Furthermore, the presence of molecular defects may be revealed using a thermal camera, instead of an optic one, to scan the foil surface, when placed on a heating pad~\cite{Constantinescu:2026gtq}.  The scanning of etched NTD sheets, often amounting to hundreds of square metres, is a laborious task  requiring tremendous time and effort. Lately, considerable effort is being invested in developing and utilising machine-learning techniques to automatically identify and classify etch-pits~\cite{Viglietti:2023,Nikaido:2024,Taguchi:2024,ML:2026}.

Concerning the material used, plastic CR-39\textsuperscript{\textregistered} is the most sensitive having a (low) threshold of $z/\beta \simeq 5$~ \cite{Cecchini:1995rw,Cecchini:2016vrw,Kalliokoski:2025sog}. The Makrofol\textsuperscript{\textregistered} and Lexan\texttrademark\ polycarbonates have a higher threshold of $z/\beta \simeq 50$, hence they are sensitive only to magnetic charges, to extremely slow HIPs and to very high electric charges.  On the downside, CR-39\textsuperscript{\textregistered} also registers background radiation, e.g.\ from neutrons, which may make etch-pit identification more difficult (cf.\ shallow circular hollow features in Fig.~\ref{fig:etchpit}). 
The efficiency of these detectors also depends on the incidence angle; the steeper the angle the lower the threshold. NTDs are calibrated by exposure to heavy-ion beam fragments~\cite{Cecchini:1995rw,Patrizii:2010jla,Kalliokoski:2025sog}. The different energy loss dependence on $\beta$ of electric~\eqref{eq:bethe-bloch} and magnetic~\eqref{eq:ahlen} charges, depicted in Fig.~\ref{fig:dedx}, affects the evolution of the etch-pit cone shape in subsequent layers of NTDs, thus providing a means to discriminate between magnetic and electric charge. 

As an alternative to NTDs, the use of solid state breakdown counters (SSBC) has been proposed~\cite{Ostrovskiy:2014hfa}. The SSBC exhibits high \dedx thresholds, convenience of electronic registration and simplicity of fabrication, operation, and signal extraction; however more research and development is required to render it an attractive means for MM detection. 

\subsection{Induction}\label{sc:squid}

\begin{figure}[htb]
\centering
\includegraphics[width=0.7\textwidth]{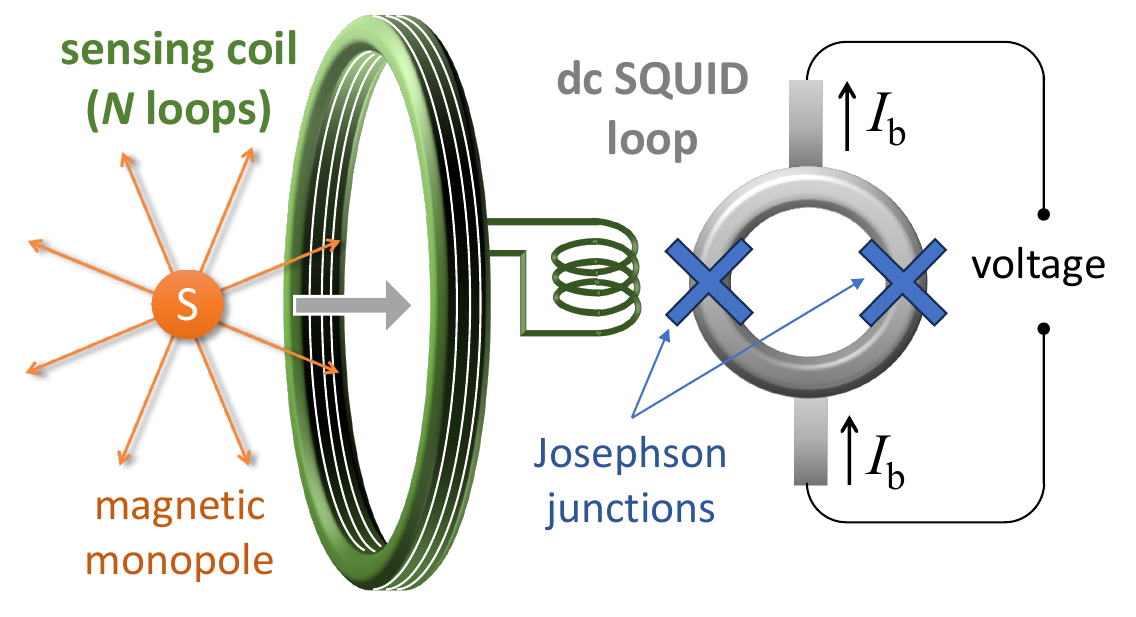}
\caption{Simplified schematic view of a dc~SQUID (not to scale). When a MM or a sample passes through the detection coil (green) coupled to a SQUID, a net supercurrent is registered in the SQUID ring. \vspace{0.2cm}}
\label{fig:squid} 
\end{figure}

The induction technique for detecting MMs is based on the long-range electromagnetic interaction of a MM with the microscopic state of a superconducting loop and it is directly sensitive to the magnetic charge $g$~\cite{Alvarez:1971zt}.  The magnetic flux of a MM of charge $g$ passing through the loop is given by $\mathrm{\Phi} = 4\pi g$, in natural units. The induced persistent electric current $\mathrm{\Delta} i$ in a coil with $N$ turns and inductance $L$ is given by the formula:
\be
\mathrm{\Delta} i = 4 \pi N g / L.
\label{eq:squid}
\ee
To measure such a minuscule magnetic flux, the pick-up coil is inductively coupled to a superconducting quantum interference device (SQUID), as illustrated in Fig.~\ref{fig:squid}. Two Josephson junctions are connected in parallel in the superconducting loop, characterising it by an elementary flux quantum $\mathrm{\Phi}_0 = 2\pi/e$, where the factor two arises from the electrons appearing as Cooper pairs~\cite{Fagaly:2006}. 

\begin{figure}[htb]
\sidecaption
\includegraphics[width=0.5\textwidth]{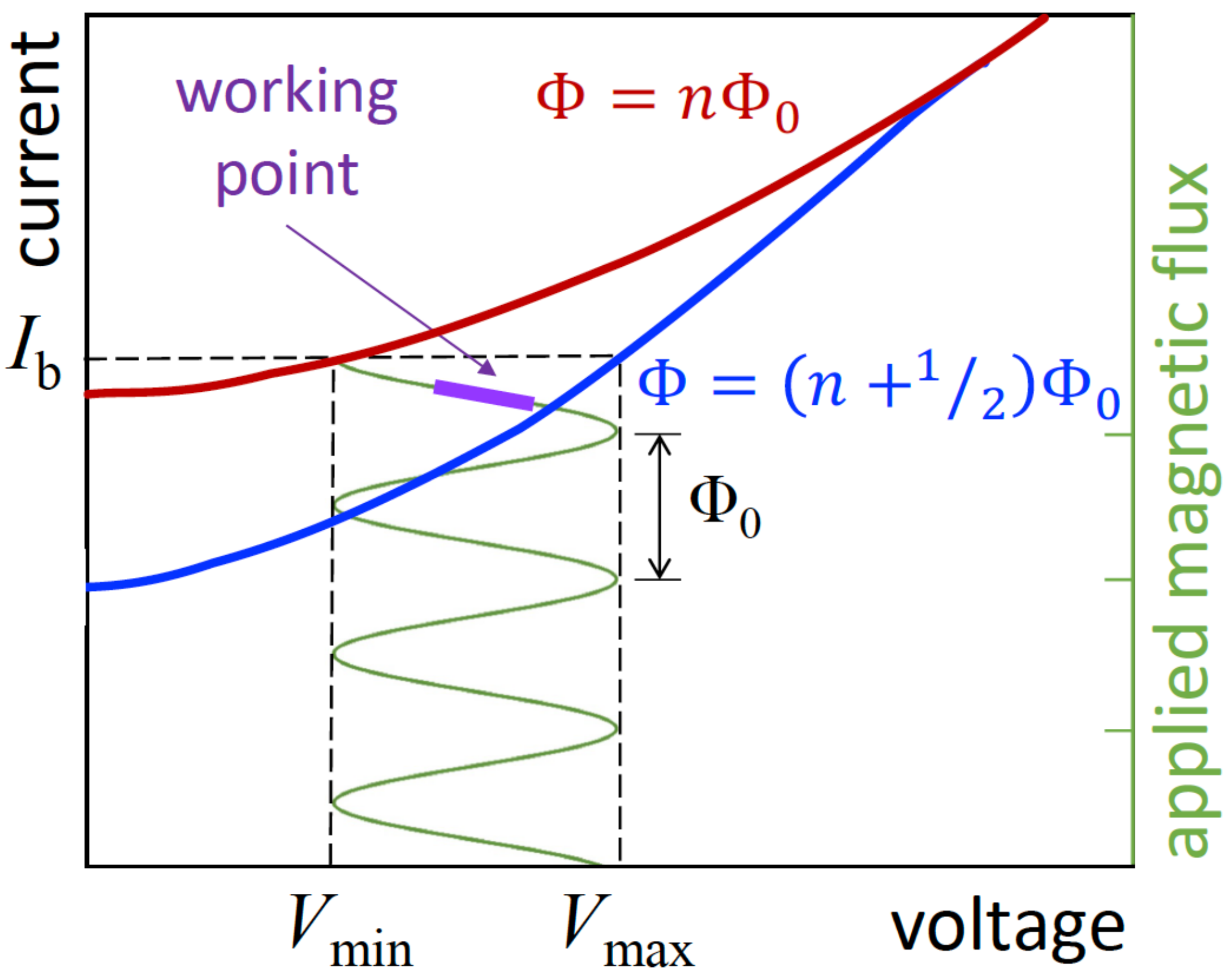}
\caption{Typical $I$--$V$ curve and SQUID transfer function (flux-to-voltage response) illustrating periodicity with applied magnetic flux.  The current value is  between the $I$--$V$ curve for integer $(n\mathrm{\Phi}_0)$ (red) and odd half-integer $((n+\hf)\mathrm{\Phi}_0)$ (blue) values of $\mathrm{\Phi}$. The working point where the voltage is almost linearly dependent on  $\mathrm{\Phi}$ is shown (thick purple). \vspace{0.2cm}}
\label{fig:squid-iv} 
\end{figure}

When a dc~SQUID is biased with a constant current, $I_\text{b}$, the temporal average voltage across the SQUID is modulated  with a periodicity of $\mathrm{\Phi}_0$,  as shown in Fig.~\ref{fig:squid-iv}. The $I$--$V$ graph displays two curves corresponding to integer $(n\mathrm{\Phi}_0)$ and odd half-integer $((n+\hf)\mathrm{\Phi}_0)$ values of the applied flux $\mathrm{\Phi}$. Therefore, for a given $I_\text{b}$,  a $V_\text{min}$ and $V_\text{max}$ are defined. The SQUID is operated on the steep part of the $V$--$\mathrm{\Phi}$ response curve, also shown in the diagram. 

The major background for these experiments are small changes in Earth's magnetic field, therefore, shielding of the ambient field is required with extreme caution, leading to high costs for detectors with broad surveillance areas. As a consequence, this technique is no longer used to search directly for MMs of cosmic origin, yet it is still widely used in searches for MMs bound in matter, as we shall discuss in Sections~\ref{sc:bound}, \ref{sc:coll-direct} and \ref{sc:moedal}. The advantage of this method is that it is fast and allows for a virtually infinite number of measurements for any sample showing signal-like behaviour.

\subsection{Cherenkov radiation}\label{sc:cherenkov}

When traversing a medium such as water or ice, relativistic MMs would lose some of their energy to Cherenkov radiation. When the MM speed exceeds the group velocity of light in that particular medium, photons are emitted from excited atoms in the medium. Electrically charged particles also give rise to Cherenkov radiation, yet the number of photons emitted is then much smaller. In water and ice, having a refractive index $n_{\mathrm r} \simeq 1.33$, a MM of one Dirac charge generates $(\gd n_{\mathrm r}/e)^2 \simeq 8,\!300$ more photons than a particle with one electric unit charge traveling with the same speed. The radiation can only be produced by particles with speeds above a threshold of $\beta_\text{thr} = 1/n_{\mathrm r} \simeq 0.76$. The photons are emitted coherently under a fixed angle $\cos\theta_{\mathrm C} = 1/\beta n_{\mathrm r}$, which for water or ice is $\theta_{\mathrm C} \simeq 41.2^\circ $ for relativistic monopoles. As demonstrated in Section~\ref{sc:nu-telescopes} below, this phenomenon is the principal technique used in neutrino telescopes to look for MMs.

\subsection{Catalysis of nucleon decay}\label{sc:callan}

\begin{figure}[htb]
\sidecaption
\includegraphics[width=0.42\textwidth]{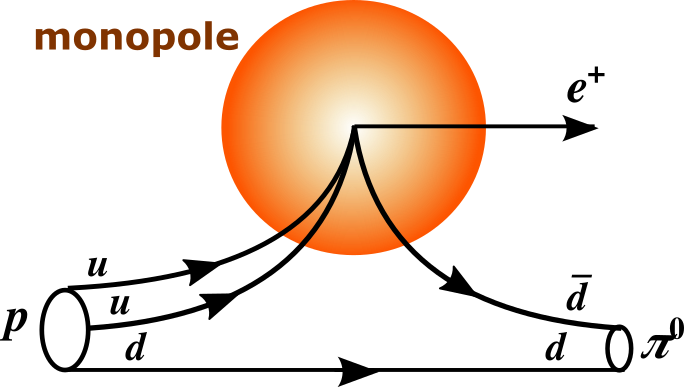}
\caption{Schematic diagram of MM catalysis of proton decay $p \rightarrow e^+ + \pi^0$, via the Callan--Rubakov mechanism, whereby baryon number violation is mediated by super heavy gauge bosons of the pertinent GUT theory in the presence of a MM. \label{fig:cr}}
\end{figure}

It has been postulated that the boson in the core of a GUT monopole may cause nucleons to decay by performing transitions between quarks and leptons; the so-called \emph{Callan--Rubakov mechanism}~\cite{Rubakov:1981rg,Rubakov:1983sy,Callan:1982ac}. Processes of this type, such as $uud \to e^+ \bar{d}d$ and $udd \to e^+ \bar{u}d$ (cf.\ Fig.~\ref{fig:cr}), violate the baryon-number conservation. The process cross section, $\sigma_0$, is of the same order as that of the strong interactions, while the branching fractions of the aforesaid transitions exceed 90\%. 

The decay products, being much lighter than their parents, are highly relativistic, e.g.\ pions subsequently decay into neutrinos. Besides this two-stage neutrino production, the possibility of utilsing \emph{direct} proton decay to \emph{monochromatic} neutrinos from the Sun to detect this effect has been proposed~\cite{Houston:2018rvz}. Therefore, along the trajectory of a catalysing MM in medium, outbursts of Cherenkov radiation would occur. The catalysis cross section depends on the monopole--nucleon relative velocity as $\sigma_\text{cat}=(\sigma_0/\beta)F(\beta)$, where $F(\beta)$ is a correction factor relevant for speeds below a threshold. Hence, nucleon-decay catalysis allows detection of extremely massive sub-relativistic MMs, e.g.\ in neutrino telescopes, as discussed in Section~\ref{sc:nucl-decay}.

\section{Searches for cosmic monopoles}\label{sc:cosmics}

\subsection{Formation and astrophysical considerations}\label{sc:formation}

If MMs of cosmic origin do exist, they must have been formed presumably as ('t Hooft--Polyakov) topological defects via the Kibble mechanism~\cite{Kibble:1976sj,Kibble:1980mv} during a phase transition in the early Universe~\cite{Preskill:1979zi,Zeldovich:1978wj,Vilenkin:2000jqa}.\footnote{It has been proposed recently the possibility for MMs to be formed \emph{locally} in Earth or interstellar space, as discussed in Section~\ref{sc:cosmic-colliders}.} The estimated relic density of such GUT monopoles leads to an overabundance, inconsistent with their non-observation,  implying an overclosure of the Universe~\cite{Kolb:1990vq}, thus constituting the \emph{monopole problem}. Cosmic inflation solves this contradiction by predicting exponential expansion that dilutes the density of these MMs, making them virtually undetectable~\cite{Linde:1981mu}. Alternative solutions involve complex patterns of symmetry breaking~\cite{Langacker:1980kd}, that may enhance MM annihilation~\cite{Perri:2025vtn},  primordial black holes~\cite{Stojkovic:2004hz,Stojkovic:2005zh}, and domain walls in the early Universe~\cite{Dvali:1997sa,Pogosian:1999zi}.  MMs can connect flavour-motivated extensions of the SM with the inflationary history of the Universe~\cite{Pesut:2026fyl}.  The Schwinger effect in strong magnetic fields may also contribute to the MM density~\cite{Kobayashi:2021des,Kobayashi:2022qpl}. 

\subsubsection{Monopoles and dark matter}\label{sc:dm}

Before moving to the cosmological and observational constraints on MMs, it is worth highlighting their connection with dark matter (DM). Point-like topological defects, such as MMs, that arise in the breaking of appropriate gauge symmetry groups, can be produced non-thermally in the early Universe via the Kibble--Zurek extended mechanism~\cite{Zurek:1985qw} and may be abundant enough to play the role of cosmological DM candidates with masses in the range 1--$10^9~\gev$~\cite{Murayama:2009nj}. This class of DM phenomenology encompasses several types of GUT-MM theories, including relatively low-mass ones~\cite{Kephart:2017esj}, global monopoles, discussed in Section~\ref{sc:global}, as well as their string-inspired MM extensions~\cite{Mavromatos:2016mnj,Mavromatos:2018drr}. 

It has been pointed out that a dark sector with MMs will also contain massless dark photons and charged massive vector bosons, leading to a combination of magnetically and electrically charged DM candidates~\cite{Khoze:2014woa}. A recent work claims that the dark matter relic density will be dominated by the lightest electrically charged particle~\cite{Brummer:2025inh}. 

Theories of two photons~\cite{Cabibbo:1962td,Salam:1966bd,Zwanziger:1970hk,Singleton:1995cc,Singleton:2011ru}, containing a typical photon interacting with electric charge and an extraordinary photon interacting only with magnetic charge\footnote{The two-photon formalism may form the basis for building effective field theories for studying MMs, e.g.\  resummation schemes (cf.~\ref{sc:resum})} may have implications for DM through the so-called \emph{dark monopoles}~\cite{Brummer:2009oul,Terning:2018lsv}. The dark and ordinary sectors interact feebly, yet  experimentally connected through a portal interaction due to the kinetic mixing between the dark and ordinary photons~\cite{Holdom:1985ag,Fabbrichesi:2020wbt,Vento:2025tql}. This kind of dark photon, the \emph{magnetic photon}, has been sought after in laboratory searches based on their easy penetration to metals, with negative results~\cite{Lakes:2004rc}. 
A dipole solution of Kaluza--Klein multidimensional theories associated with a monopole--antimonopole bound state, the so-called \emph{Kaluza--Klein monopolium},\footnote{Its decaying counterpart in gauge theories is discussed in Section~\ref{sc:monopolium}.} is a viable candidate for a primordial dark matter constituent, as it is classically stable and,  therefore, very long lived~\cite{Vento:2020vsq}.

MMs may live exclusively in the hidden sector of a theory involving extra space-like dimensions with interesting observational and experimental implications~\cite{Mitsou:2013rwa,Mitsou:2019xzu} as DM candidates through appropriate portal interactions that connect the dark and visible sectors~\cite{Lazarides:2000em,Zhang:2019ona,Daido:2019tbm,Sato:2018nqy,Baek:2013dwa,Sousa:2009is,Rahaman:2006kw,Terning:2018lsv}. A note is due here on scenarios with kinetic mixing between the visible and a hidden sector, in which in general the kinetic mixing leads to \emph{milli-magnetically charged particles}~\cite{Brummer:2009cs,Brummer:2009oul,Sanchez:2011mf}. Model-independent bounds on such particles have been set based on magnetars~\cite{Hook:2017vyc} and on the survival of galactic magnetic fields~\cite{Iguro:2024oml,Graesser:2021vkr,Graesser:2026zzs}.

\subsubsection{Interactions with magnetic fields}\label{sc:parker}

\begin{figure}[htb]
\sidecaption
\includegraphics[width=0.55\textwidth]{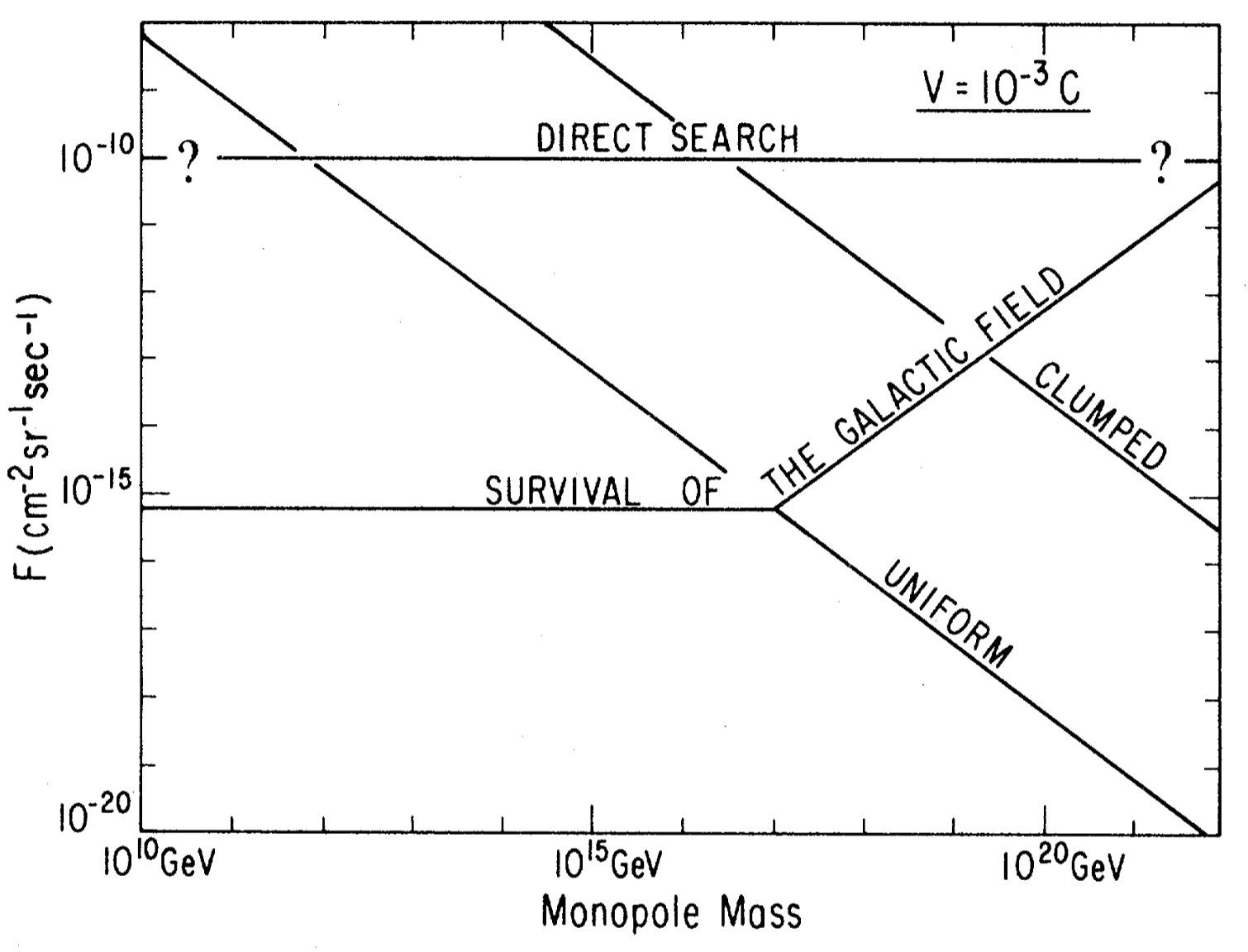}
\caption{Historical summary of MM flux limits as a function of MM mass for an initial MM velocity of $10^{-3}$, valid for uniform or isotropic velocity distribution. The (now obsolete) lines marked ``uniform'' and ``clumped'' are associated with the overclosure of the Universe discussed at the beginning of Section~\ref{sc:formation}. The ``direct search'' limit, applicable for $3\times10^{-4} \lesssim \beta \lesssim 2\times10^{-2}$, are superseded by more stringent ones. From Ref.~\cite{Turner:1982ag}. \vspace{0.2cm}}
\label{fig:parker} 
\end{figure}

The existence of the galactic magnetic field would accelerate MMs, thus draining its energy. In order for the galactic field to sustain, its dissipation must not exceed its regeneration, implying that an upper  limit on cosmic-MM flux, $\mathrm{\Phi}_\text{cosm}$, should be respected, the  \emph{Parker bound}~\cite{Parker:1970xv}.
A reexamination of this limit considering the observed structure of the galactic magnetic field, showed that it depends on the MM mass and velocity, leading to the so-called \emph{Turner--Parker--Bogdan (TPB) bound}~\cite{Turner:1982ag}, shown in Fig.~\ref{fig:parker} from the original paper. By assuming reasonable values of  pertinent astrophysical parameters, the Parker bound lowers to
\be
\mathrm{\Phi}_\text{cosm} \lesssim \begin{cases}
			10^{-15}~\cmssr, & M  \lesssim 10^{17}~\gev \\
            		10^{-15} \left( \frac{M}{10^{17}~\mathrm{GeV}} \right)~\cmssr, & M  \gtrsim 10^{17}~\gev
		 \end{cases},
\label{eq:parker}
\ee
where $M$ is the MM mass; a magnetic charge of 1\gd is assumed. A MM mass of $\sim 10^{17}~\gev$ separates the two regimes with respect to the value $\beta_\text{mag}$ for which a MM initially at rest would accelerate through a region of coherent field: (i) $\beta \lesssim \beta_\text{mag}$, where MMs are easily deflected by the magnetic forces, and (ii) $\beta \gtrsim \beta_\text{mag}$, where the magnetic force is a small perturbation to the MM motion. Even tighter bounds, such as the \emph{extended Parker bound}, can be set considering galactic seed field and progenitors, yet with larger uncertainties~\cite{Adams:1993fj}. 

Likewise, bounds on MM flux are imposed by the survival of \emph{primordial} magnetic fields~\cite{Long:2015cza,Kobayashi:2022qpl,Papanikolaou:2023nkx,Kobayashi:2023ryr}. The MM production and acceleration in magnetic fields is being studied in depth recently~\cite{Kobayashi:2022qpl,Kobayashi:2023ryr,Perri:2023ncd,Zhang:2026kgb}.  
A mechanism for the MM production from interactions of cosmic rays and interstellar medium, leads to generalised Parker-like bounds~\cite{Iguro:2024oml}. 
Further constraints from astrophysical objects come from neutron stars~\cite{Giacomelli:1984gq,Gould:2017zwi,Hook:2017vyc}. More details on MM production and acceleration mechanisms and their implications for terrestrial experiments are presented in Section~\ref{sc:cosmic-colliders}.

Monopoles of cosmic origin~\cite{Rajantie:2016paj,Patrizii:2015uea,Patrizii:2019eud,Spurio:2019oaq,Mavromatos:2020gwk} can be detected by exploiting any of the techniques outlined in Section~\ref{sc:techniques}. The search may involve  material where MMs may be either trapped or may have left tracks, as discussed in Sections~\ref{sc:bound} and~\ref{sc:mica}, respectively. However, most of the searches, described in Section~\ref{sc:direct}, seek monopoles in-flight interacting with detectors covering a wide spectrum of MM masses and velocities. Finally, searches for effects of MM-catalysed decays, e.g.\ proton decay, are discussed in Section~\ref{sc:nucl-decay}. 

\subsection{Monopoles trapped in matter}\label{sc:bound}

\begin{figure}[htb]
\sidecaption
    \includegraphics[width=0.55\linewidth]{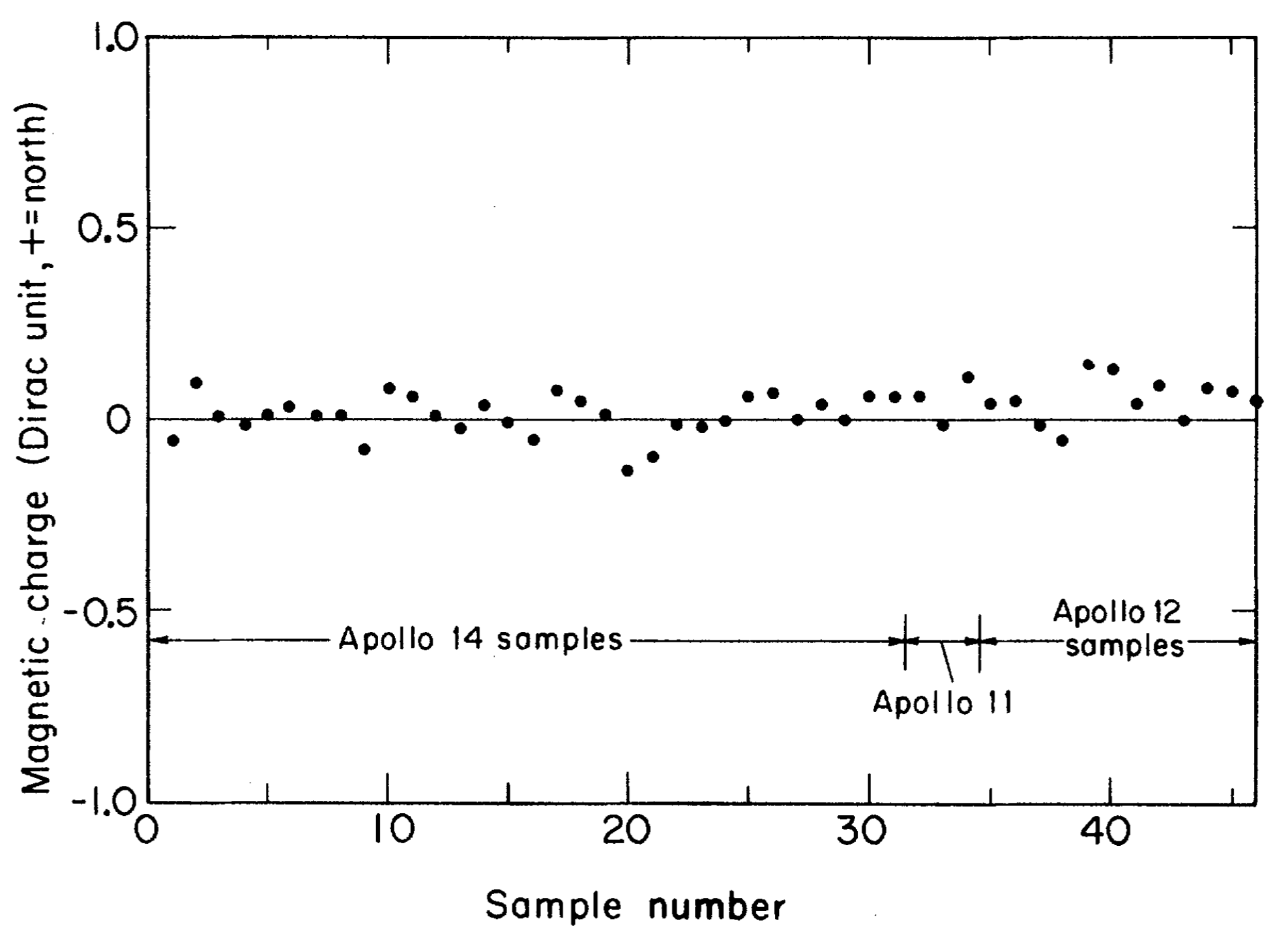}
    \caption{Magnetic-charge SQUID measurements obtained by scanning lunar-rock samples collected during Apollo expeditions to the Moon. From Ref.~\cite{Ross:1973it}. \vspace{0.2cm}}
    \label{fig:lunar}
\end{figure}

During the last decades, the induction technique has been mostly deployed in searches for MMs \emph{bound in matter}, such as lunar rocks~\cite{Eberhard:1971re,Ross:1973it} (cf.\ Fig.~\ref{fig:nova} in Section~\ref{sc:slow-mms}), meteorites~\cite{Kovalik:1986zz,Jeon:1995rf}, seawater~\cite{Kovalik:1986zz}, iron ores~\cite{Ebisu:1986dw} and ferromanganese nodules~\cite{Kovalik:1986zz}, by passing samples through superconducting loops.  The presence of a MM captured in the material would manifest itself as an alternation in the persistent-current value before and after the passage of the sample through the sensing coil, as discussed in Section~\ref{sc:squid}. 
The (calibrated to magnetic charge) SQUID measurements from Apollo-collected lunar rocks shown in Fig~\ref{fig:lunar}, were compatible with a zero magnetic charge, hence MM-flux bounds, incorporated in Fig.~\ref{fig:nova}, were imposed~\cite{Ross:1973it}. 
With this method, a stringent upper limit on the MMs per nucleon ratio of $\sim10^{-29}$ has been obtained~\cite{Kovalik:1986zz,Jeon:1995rf}. Moreover, as we shall see in Section~\ref{sc:colliders}, SQUIDs have been used in experiments to look for MMs produced in high-energy collisions and trapped in volumes surrounding the interaction points.

\begin{figure}[htb]
\sidecaption
    \includegraphics[width=0.4\linewidth]{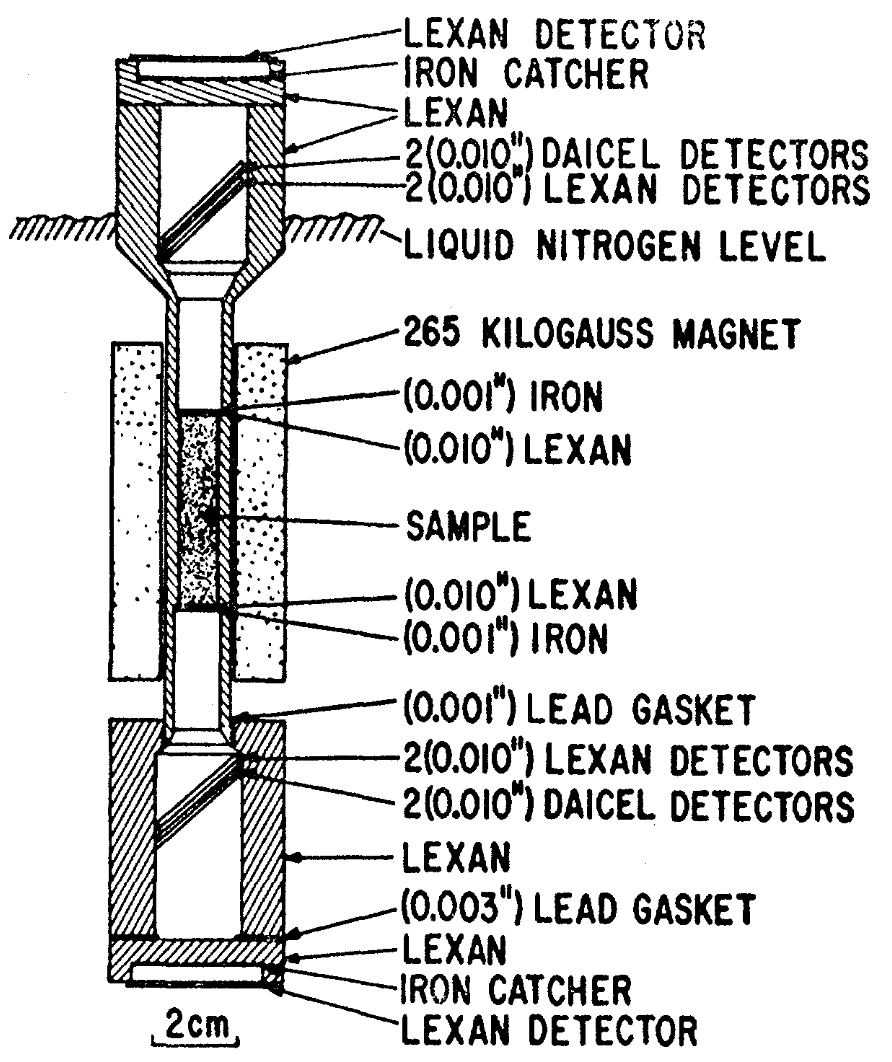}
    \caption{Device for extracting MMs from a sample, passing them through a detection system, and collecting them. The sample is placed in the centre of the magnet coil. During the magnetic-field rise time, the MM will migrate upward/downward through a Lexan\texttrademark\ sheet into a ``retarder''  sheet of iron. There, it will be trapped until a critical field value is reached, at which time the MM is extracted from the iron and accelerated into the detection system: two Lexan\texttrademark\ polycarbonate sheets followed by two DAICEL cellulose nitrate sheets. From Ref.~\cite{Fleischer:1969mj}. \vspace{0.1cm}}
    \label{fig:extraction}
\end{figure}

Another widely used, yet less known nowadays, method is the \emph{extraction} technique, which involves the application of strong ($\geq 5~{\rm T}$) magnetic field to samples, sufficient to dislodge the bound MMs, accelerate them to an appropriate velocity and identify them via the high \dedx loss in a scintillator or an NTD array. A schematic of a typical experimental apparatus and methodology description is presented in Fig.~\ref{fig:extraction}. It has been used in the past to search for MMs in seawater, air and ocean bottom samples~\cite{Fleischer:1969mj,Fleischer:1970zy,Kolm:1971xb,Carrigan:1975bk}, as well as material exposed in high-energy collisions  (cf.\ Section~\ref{sc:tevatron}). A detailed account of searches for MMs bound in matter and a discussion on the MM binding and detachment conditions is given in Ref.~\cite{Burdin:2014xma}.

Last but not least, the Gauss's law for magnetism can be combined with precision magnetic-field measurements in a closed surface to reveal a non-zero net magnetic charge in the volume. Such an analysis was  performed in 2021 for the surface of Earth utilising data from the Swarm mission, consisting of three satellites in low-Earth near-polar orbits measuring the vector magnetic field. Bounds on the monopole moment were set at the level of $\mathcal{O}(\mathrm{nT})$~\cite{Bai:2021ewf}, which are sufficiently sensitive to constrain the local monopole energy density due to magnetic black holes~\cite{Lee:1991qs,Maldacena:2020skw}. Prospects for pursuing this line of searches in other astronomical objects, e.g.\ Mars, exist.

\subsection{Traces in matter from past MM impacts with Earth}\label{sc:mica}

Another class of experiments relied on ionisation to look for traces of traversing MMs in ancient mica ($4.6\times10^8~\mathrm{yr}$). The analysis involves a search for defects in the molecular structure of the material caused by the propagation of a MM following a similar approach as for plastic NTDs. The observed absence of MM-induced tracks in the mica detector placed an upper limit of $10^{-17}$ to $10^{-16}~\cmssr$ on the flux of supermassive MMs having velocity $\beta \sim 3\times10^{-4} - 1.5\times10^{-3}$~\cite{Price:1983ax}. This was the first direct search for MMs with adequate sensitivity to detect a flux as small as the Parker limit, followed by another that lowered slightly this limit~\cite{Ghosh:1990ki}, which is shown in Fig.~\ref{fig:super-kamio} along with other constraints in Section~\ref{sc:nucl-decay}.

\subsection{Direct cosmic searches}\label{sc:direct}

Searches for non-relativistic MMs~\cite{Patrizii:2015uea,Patrizii:2019eud,Spurio:2019oaq} have been performed on underground, surface and balloon-borne experiments targeting GUT MMs spanning masses of 100--$10^4~\tev$ with a velocity range of $10^{-5} < \beta < 1$. When considering down-going MMs, the main background sources are cosmic-ray muons and natural radioactivity, which can be partly rejected by two classes of detection methods, also applicable to heavy stable electrically charged particles (cf.\ Section~\ref{sc:hecos}). The first method, suitable for very slow monopoles, is based on a time-of-flight measurement or a wide signal in thick detector plates. The second method is complementary to the first and provides sensitivity for less massive and faster monopoles and involves anomalously large energy loss. To date, there is no experimental evidence for cosmic MMs, only bounds on their flux as a function of velocity and the mass more recently~\cite{Iguro:2021xsu,Iguro:2024oml,Perri:2025qpg}.  The present limits for slow-moving MMs are discussed in Section~\ref{sc:slow-mms}, while searches for fast and ultra-fast MMs through Cherenkov radiation are presented in Section~\ref{sc:nu-telescopes}.

\subsubsection{Slow monopoles}\label{sc:slow-mms}

\begin{figure}[ht]
  \centering
  \includegraphics[width=0.85\linewidth]{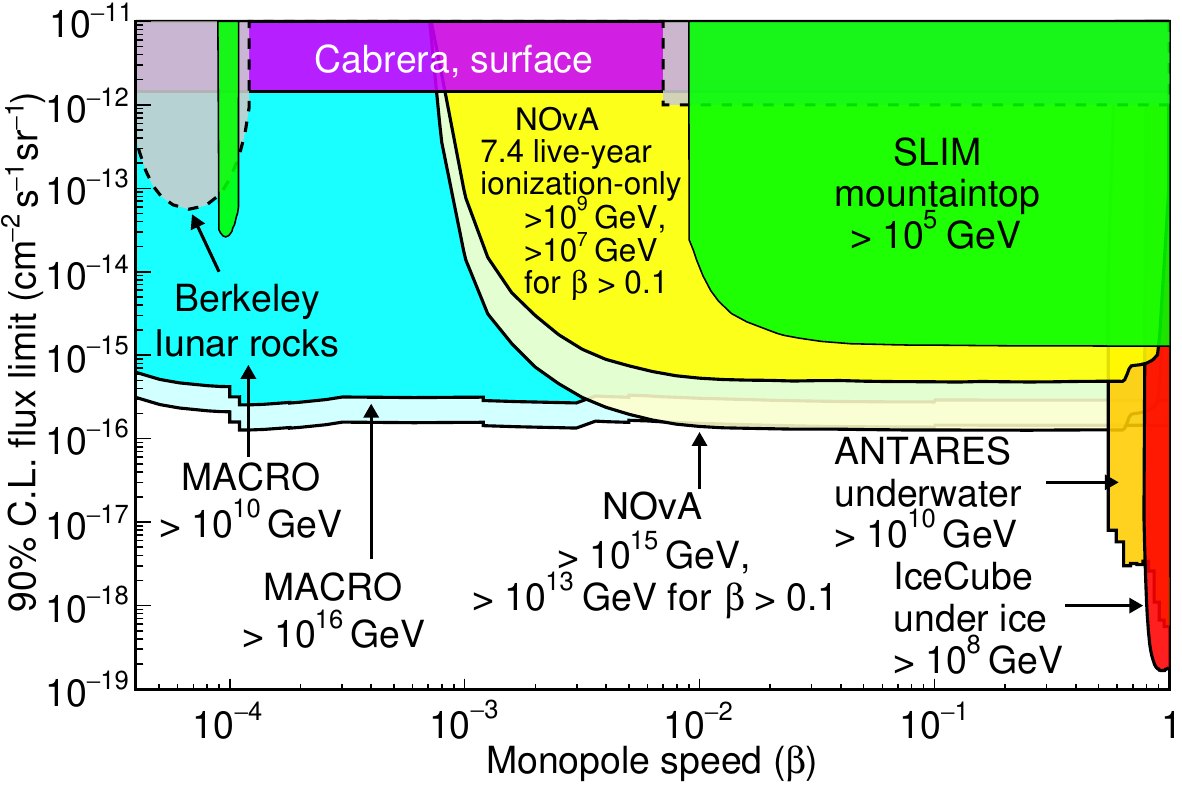}
  \caption{90\%-CL upper limits on GUT MM flux versus velocity $\beta$ for a magnetic charge of $g = \gd$ for several experiments: three-loop SQUID~\cite{Cabrera:1983iie}, SLIM~\cite{Balestra:2008ps}, MACRO~\cite{MACRO:1993oqw}, lunar rocks~\cite{Eberhard:1971re}, ANTARES~\cite{ANTARES:2022zbr,ANTARES:2025ojl}, IceCube~\cite{IceCube:2021eye}, and \nova~\cite{NOvA:2025eow}. In each region, the experiment with the strongest mass limit (least overburden) is shown above the others. Detailed results for fast MMs from neutrino telescopes are given in Fig.~\ref{fig:antares}. Adapted from Ref.~\cite{NOvA:2025eow}.}
\label{fig:nova}
\end{figure}

The Monopole, Astrophysics and Cosmic Ray Observatory (MACRO)~\cite{MACRO:2002kki} was a large underground detector operated in the Gran Sasso laboratory during the 1990s at a more than 3100~m of water-equivalent depth. It provided the best limits for super-heavy GUT MMs up to date with a sensitivity that covers most of the phase space in Fig.~\ref{fig:nova}, mostly thanks to the redundancy and complementarity of its various detector components: liquid scintillation counters; limited streamer tubes; and NTDs~\cite{MACRO:1993oqw,MACRO:1997pqq,MACRO:2001vyq,MACRO:2002jdv}. However, due to its underground location, it was not sensitive to low-energy MMs, which are blocked by the Earth~\cite{Derkaoui:1998uv}. As shown in Fig.~\ref{fig:nova}, the upper limit on the local monopole flux set is $1.4\times 10^{-16}~\cmssr$ at 90\% confidence limit (CL),  i.e.\ well below the Parker bound, in almost all the $\beta$ range for GUT MMs~\cite{MACRO:2002jdv}. 

Besides some (early) sea-level NTD experiments~\cite{Fleischer:1971bd,Bartlett:1981hy,Doke:1983if}, underground experiments include the Ohya stone quarries near Tokyo, which used a 2000~$\mathrm{m}^2$ array of CR-39\textsuperscript{\textregistered} NTDs to place an upper flux limit of $3.2\times 10^{-16}~\cmssr$ on MMs in the velocity range of $4\times 10^{-5} < \beta <1$~\cite{Orito:1990ny}. Another telescope, deployed at Baksan~\cite{Novoseltsev:2006mw} in Russia, used liquid scintillation counters to probe both slow (Baksan1) and fast (Baksan2) MMs. Soudan~2~\cite{Soudan-2:1992olq} in the United States, on the other hand, was a large fine-grained tracking calorimeter composed of long drift tubes. Similar techniques were used earlier by other nuclear-decay detectors, namely, the tracking calorimeter Kolar Gold Fields (KGF)~\cite{Krishnaswamy:1984fu}, deployed in India, and the Mont-Blanc Nucleon Stability Experiment~\cite{Battistoni:1983ka}, which used plastic streamer tubes; both detectors placed looser flux bounds than MACRO, Ohya and Baksan. Bounds set by Baksan, KGF and Soudan~2 are depicted in Fig.~\ref{fig:super-kamio} along with other experimental results.

The SLIM detector, installed at high altitude at the Mt Chacaltaya laboratory in Bolivia with an elevation of $5400~\mathrm{m}$, probed intermediate-mass MMs ($10^5 \lesssim M \lesssim 10^{12}~\gev$), well below the GUT scale, which do not have enough energy to penetrate the entire atmosphere. The SLIM NTDs array covered an area $427~{\mathrm m}^2$ that after four years of exposure did not observe any signal of MMs and set the limits shown in Fig.~\ref{fig:nova} for single magnetic charge~\cite{Balestra:2008ps}. This detector was also sensitive to MMs of charge 2\gd in the range $4\times 10^{-5} < \beta < 1$.

\nova is a long-baseline neutrino experiment studying neutrino oscillations in the Fermilab NuMI beam. The far detector, located at the surface, consists of liquid scintillator cells read out at both ends by avalanche photodiodes. This detector covers a previously unexplored phase-space region of intermediate-mass slow MMs due to its location on the surface. After a three-month exposure, an upper flux limit of $2\times 10^{-14}~\cmssr$ at 90\% CL for MM velocity $6\times 10^{-4} < \beta < 5\times 10^{-3}$  and mass greater than $5 \times 10^{8}~\gev$~\cite{NOvA:2020qpg}. Recently, \nova reported a more stringent 90\% CL limit of \mbox{$8 \times 10^{-16}~\cmssr$} (shown in Fig.~\ref{fig:nova}), in the range \mbox{$0.005 < \beta < 0.8$}, corresponding to MMs with masses greater than $10^{9}~\gev$ for all velocities, and greater than $10^{7}~\gev$ for $\beta > 0.1$~\cite{NOvA:2025eow}. Using timing information, this limit has recently been extended to the lower-mass region~\cite{NOvA:2026pts}.

\subsubsection{Cherenkov telescopes and relativistic monopoles}\label{sc:nu-telescopes}

\begin{figure}[ht]
\centering
\includegraphics[width=0.8\linewidth]{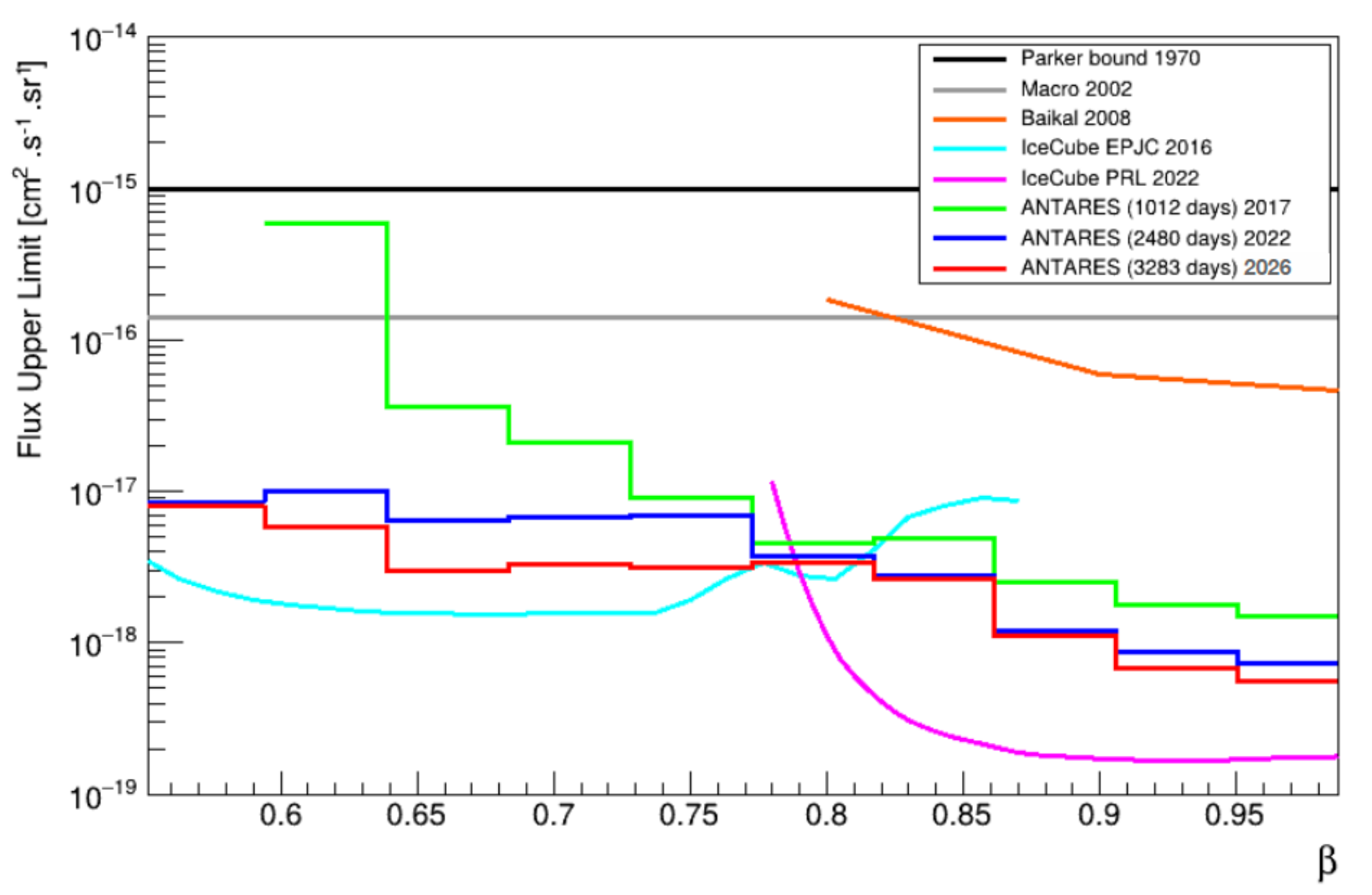}
\caption{Summary on 90\% CL singly charged monopole flux upper limits from ANTARES 14-yr (red)~\cite{ANTARES:2025ojl}, 
ANTARES 10-yr (blue)~\cite{ANTARES:2022zbr}, ANTARES 5-yr (green)~\cite{ANTARES:2017qjw}, IceCube 8-yr (magenta)~\cite{IceCube:2021eye}, IceCube 1-yr (cyan)~\cite{IceCube:2015agw}, Baikal (orange)~\cite{BAIKAL:2007kno} and MACRO (gray)~\cite{MACRO:2002jdv}, as well as the Parker bound (black)~\cite{Parker:1970xv}. Adapted from Ref.~\cite{ANTARES:2025ojl}.}
\label{fig:antares}
\end{figure}

Relativistic monopoles can be sought through the emittance of Cherenkov radiation, when traveling through a homogeneous and transparent medium, such as ice or water, which can be detected by neutrino telescopes, which feature arrays or strings of photomultiplier tubes. Neutrino telescopes such as Baikal~\cite{BAIKAL:2007kno}, AMANDA~\cite{Abbasi:2010zz}, ANTARES~\cite{ANTARES:2011rzb,ANTARES:2017qjw,ANTARES:2022zbr,ANTARES:2025ojl}, and IceCube~\cite{IceCube:2012khj,IceCube:2014xnp,IceCube:2015agw,IceCube:2021eye} were/are sensitive to the huge quantity of visible Cherenkov light emitted by a MM with $\beta > 0.76$ (\emph{direct} Cherenkov). Additional light is produced by Cherenkov radiation from $\delta$-ray electrons along the MM path for velocities down to $\beta = 0.625$  (\emph{indirect} Cherenkov). Furthermore, luminescence may be induced by molecular excitation of the medium for MM velocities of $\beta > 0.01$~\cite{ObertackePollmann:2016uvi}. The second phase of the Antarctic Muon And Neutrino Detector Array (AMANDA-II), the predecessor of IceCube, set an upper limit on $\beta=1$ MMs flux of $3.8 \times 10^{-17}~\cmssr$~\cite{Abbasi:2010zz}. Latest results for relativistic MMs from ANTARES~\cite{ANTARES:2025ojl} and IceCube~\cite{IceCube:2021eye} are depicted in Fig.~\ref{fig:nova} and in more detail in Fig.~\ref{fig:antares}, together with previous results of these and other experiments. The strongest upper flux bounds were set by IceCube for $ 0.80 \lesssim \beta \lesssim 0.995$, for MM masses above $10^{8}$--$10^{10}~\gev$ depending on the zenith angle~\cite{IceCube:2021eye}. As is the case for neutrinos, a large background from cosmic muons inhibits searches for down-going candidates; up-going MMs having traversed the Earth before reaching the detector are probed instead. It is worth noting that GUT supermassive MMs are unlikely to reach (nearly) relativistic velocities.

\begin{figure}[htb]
\sidecaption
    \includegraphics[width=0.55\linewidth]{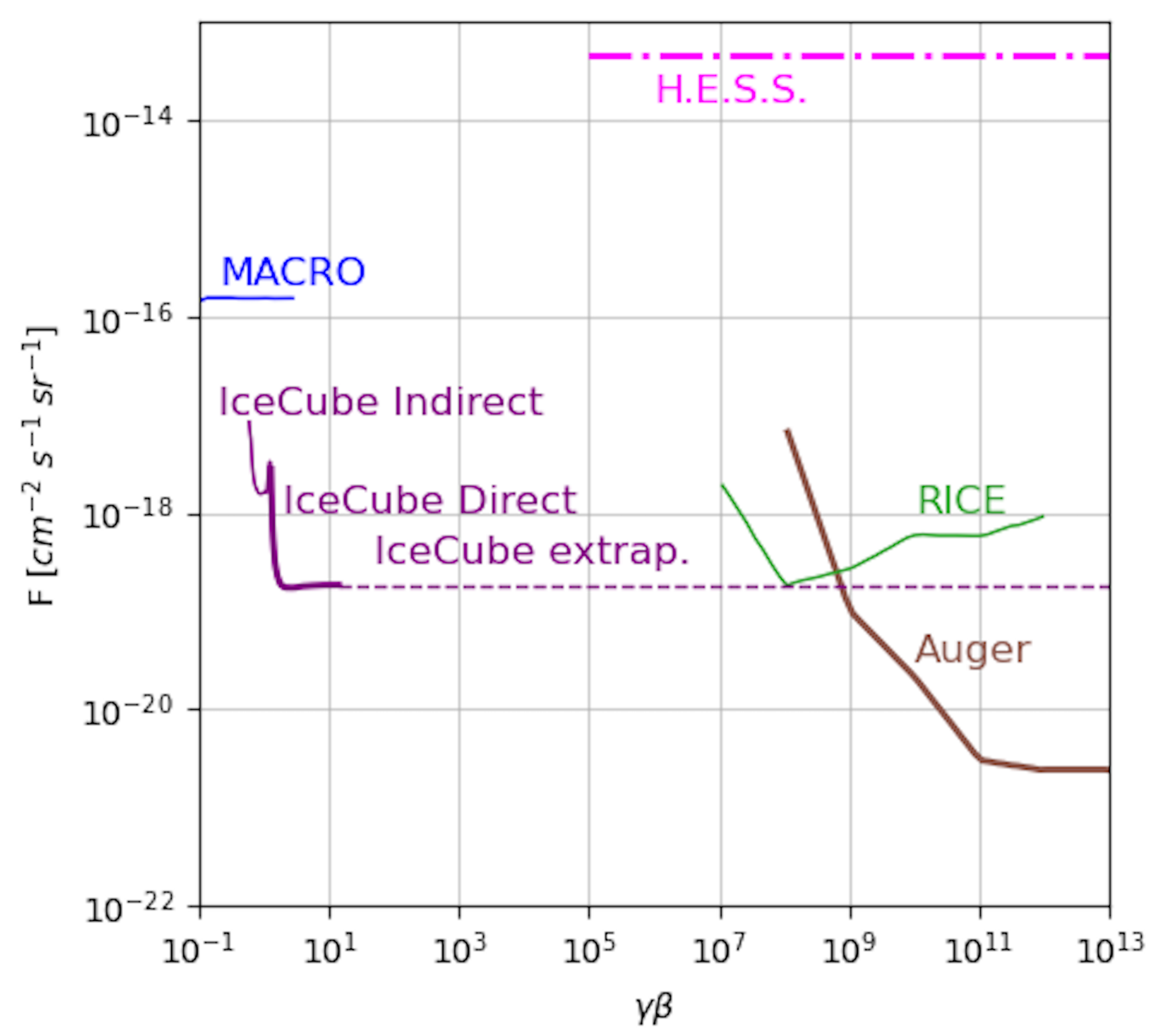}
    \caption{Experimental upper limits at 90\% CL on  $1\gd$-MM flux in terms of the MM velocity $\beta$ multiplied by the Lorentz factor $\gamma$.   Shown are limits from MACRO~\cite{MACRO:2002kki} (blue), IceCube~\cite{IceCube:2021eye} (purple), RICE~\cite{Hogan:2008sx} (green), Auger~\cite{PierreAuger:2016imq} (brown), and a preliminary result for H.E.S.S.~\cite{Spengler:2009} (magenta). From Ref.~\cite{Perri:2025qpg}. \vspace{0.4cm}
}
    \label{fig:auger-rice-hess}
\end{figure}

The flux of ultra-relativistic monopoles has been constrained by the Pierre Auger Observatory, which was sensitive to MMs with Lorentz factor values \mbox{$\gamma \sim 10^9$--$10^{12}$}, leading to flux limits in the range $2.5 \times 10^{-21} - 10^{-15}~\cmssr$~\cite{PierreAuger:2016imq}, shown in Fig.~\ref{fig:auger-rice-hess}. Two other experiments exploited the radio-wave pulses from the interactions of a primary particle with ice to search for MMs. The Radio Ice Cherenkov Experiment (RICE), consisting of radio antennas buried in the Antarctic ice, set a flux upper limit of $10^{-18}~\cmssr$ at 95\% CL for intermediate-mass MMs with $10^7 < \gamma < 10^{12}$ and a total energy of $10^{16}~\gev$~\cite{Hogan:2008sx}, also drawn in Fig.~\ref{fig:auger-rice-hess}. The ANITA-II balloon-borne radio interferometer, on the other hand, set a 90\%-CL flux upper limit on the order of $10^{-19}~\cmssr$ for a Lorentz factor $\gamma > 10^{10}$ at a total energy of $10^{16}~\gev$~\cite{ANITA-II:2010jck}.

Imaging Atmospheric Cherenkov Telescopes (IACTs) are large parabolas designed to detect Cherenkov light produced by primary $\gamma$- and cosmic rays in the Earth’s atmosphere. For ultra-relativistic MMs, this radiation would be extremely bright and visible along the entire MM trajectory. A study of the detection of MM-induced Cherenkov light by the High Energy Stereoscopic System (H.E.S.S.), an IACT array located in Namibia, showed promising sensitivity~\cite{Spengler:2011vth}. A preliminary analysis of five years of H.E.S.S.\ data resulted in a modest flux limit of $4.5 \times 10^{-14}~\cmssr$~\cite{Spengler:2009}, depicted in Fig.~\ref{fig:auger-rice-hess}.  
The Major Atmospheric Cherenkov Experiment (MACE)~\cite{Singh:2025qpi}, a single large-sized IACT operational since 2001 may render supplementary leverage in the hunt for MMs~\cite{MACE:2024} due to its location at an altitude of 4,300~m in the Himalayan desert.  
The main limitation of IACTs is their relatively narrow fields of view, an aspect expected to be improved in the coming years when the Cherenkov Telescope Array Observatory (CTA)~\cite{CTAConsortium:2017dvg} will start analysing data. This factor, together with the significantly longer operation time imply a net improvement in flux sensitivity by a factor of $\sim200$~\cite{CTAConsortium:2012fwj,Perri:2025qpg}, still inferior to the one achieved by Auger and IceCube. 

\subsection{Searches through catalysis of nucleon decay}\label{sc:nucl-decay}

\begin{figure}[htb]
\centering
\includegraphics[width=0.65\textwidth]{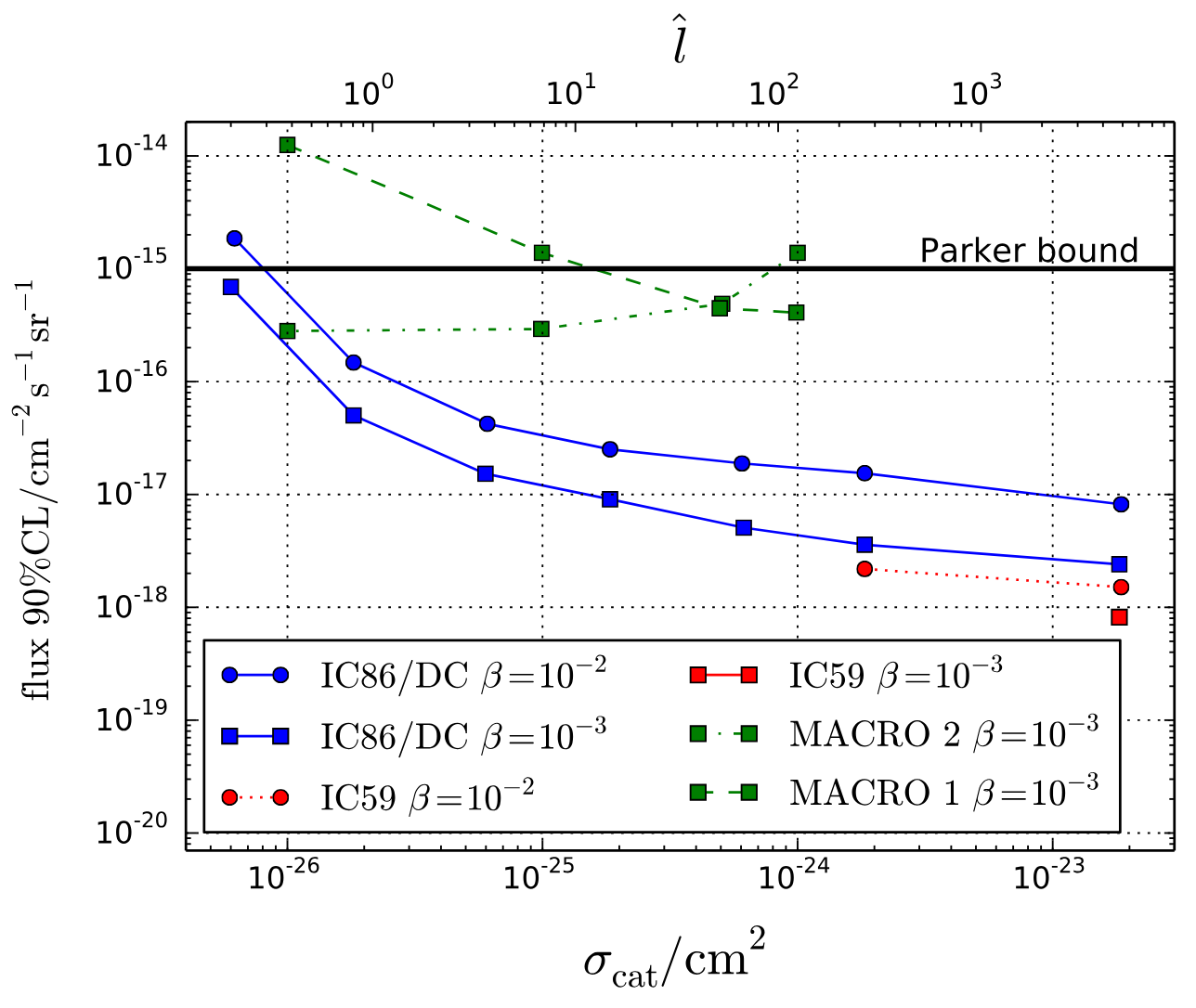}
\caption{Upper limits on the flux of non-relativistic MMs depending on the speed $\beta$ and catalysis cross section $\sigma_\text{cat}$ from the IC-59 (red) and IC-86/DeepCore (blue) analysis from IceCube~\cite{IceCube:2014xnp}. The dashed lines are limits published by  MACRO~\cite{MACRO:2002iaq}. MACRO~1 is an analysis developed for MMs catalysing the proton decay. MACRO~2 is the standard MACRO analysis, sensitive to MMs ionising the surrounding matter. In addition, the IceCube limits are shown as a function of $\hat{l}$ which is proportional to the averaged Cherenkov photon yield per nucleon decay (not valid for MACRO limits). From Ref.~\cite{IceCube:2014xnp}. \vspace{0.2cm}}
\label{fig:catalysis} 
\end{figure}

Signals of a monopole-induced decay of a nucleon, as predicted by the Callan--Rubakov mechanism~\cite{Rubakov:1981rg,Rubakov:1983sy,Callan:1982ac} and discussed earlier in Section~\ref{sc:callan}, have been sought, which, however, are sensitive to the assumed value of the catalysed-decay cross section $\sigma_{\text{cat}}$. These signals are ideal for probing ultra-heavy sub-relativistic MMs. Searches have been made with the Soudan~\cite{Bartelt:1986cv} and MACRO~\cite{MACRO:2002iaq} experiments, using tracking detectors, following nuclear-decay detectors, such as KGF~\cite{Krishnaswamy:1984fu} and the Mont-Blanc experiment~\cite{Battistoni:1983ka} (cf.\ Section~\ref{sc:slow-mms}).
Searches at the Irvine--Michigan--Brookhaven detector (IMB)~\cite{Becker-Szendy:1994kqw}, the underwater Lake Baikal experiment~\cite{Baikal:1997kuo}, and the IceCube experiment~\cite{IceCube:2014xnp} have also been performed. The resulting ($\beta$-dependent) flux limits from these experiments typically lie in the range $10^{-18}-10^{-14}~\cmssr$, as presented in Fig.~\ref{fig:catalysis} as a function of $\sigma_{\text{cat}}$. 

\begin{figure}[htb]
\sidecaption
\includegraphics[width=0.45\textwidth]{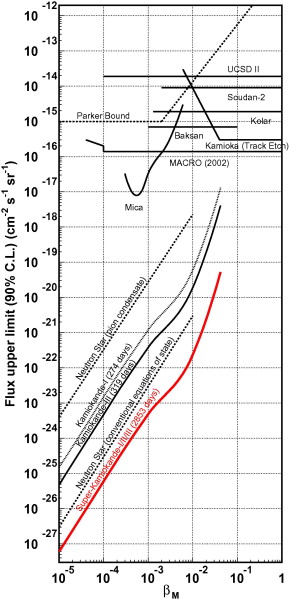}
\caption{Super-Kamiokande-obtained 90\% CL upper limits on the MM flux as a function of MM velocity, $\beta_\mathrm{M}$~\cite{Super-Kamiokande:2012tld}. Also shown are flux limits obtained by various direct detection experiments: Kamioka track etch~\cite{Orito:1990ny}, Mica~\cite{Ghosh:1990ki}, and the MACRO final result~\cite{MACRO:2002jdv}, as well as ones from indirect observations: the Parker bound~\cite{Parker:1970xv,Turner:1982ag}, limits from neutron star observations~\cite{Kolb:1984yw}, and the Kamiokande experimental results. A catalysis cross section, $\sigma_\text{cat} = 1~\mathrm{mb}$ and a MM mass of $10^{16}~\gev$ are assumed for indirect searches. From Ref.~\cite{Super-Kamiokande:2012tld}. \vspace{0.2cm}}
\label{fig:super-kamio} 
\end{figure}

A search for low-energy neutrinos, assumed to be produced from MM-induced proton decay in the Sun, was made by a 50,000-metric-ton water Cherenkov detector, the Super-Kamiokande (SK). Protons, which initially decay into pions, will ultimately produce $\nu_e$, $\nu_\mu$ and $\bar{\nu}_\mu$. After undergoing neutrino oscillation, all neutrino species appear when they arrive at the Earth, and can be detected by SK. The analysis  obtained a model- and $\beta$-dependent limit of \mbox{$6.3\times10^{-24}(0.001\beta)^2~\cmssr$}~\cite{Super-Kamiokande:2012tld}. This result, the world’s most stringent upper limit for $\beta<10^{-2}$, together with prior SK bounds and limits from other experiments is presented in Fig.~\ref{fig:super-kamio}.

Additional constraints~\cite{Giacomelli:1984gq} can be obtained by considering the effect of nucleon decay catalysed by GUT MMs on the surface heat flow in Earth~\cite{Turner:1983hu} and the Jovian planets~\cite{Arafune:1983tr}. By invoking observational limits on the diffuse ultraviolet and X-ray flux from neutron stars~\cite{Kolb:1982si,Dimopoulos:1982cz,Kolb:1984yw} and  old pulsars~\cite{Freese:1983hz}, bounds are placed on the product of the galactic flux of massive MMs and the cross section for MM-catalysed nucleon decay. 

\subsection{Evidence for monopoles?}\label{sc:cabrera}

Blas Cabrera Navarro and collaborators at Stanford University set up an experiment with a four-turn, \mbox{5-cm}-diameter loop, with its axis vertically oriented, connected to the superconducting input coil of a SQUID~\cite{Cabrera:1982gz}. He reported a single candidate event during the 151~days of its operation, on February $14^\mathrm{th}$ 1982. Although the event had the right flux-step size for a Dirac MM, another plausible explanation could be a mechanically induced offset. If this candidate event is considered to be spurious, these data set an upper limit of $6.1\times 10^{-10}~\cmssr$, which is much larger than the Parker bound. Despite further improvements of the experimental setup to suppress possible background sources, e.g.\ by using a three-loop superconducting device~\cite{Cabrera:1983iie}, this result was not confirmed. The best bound given by an induction detector on cosmic MMs was obtained later by the same group: 90\% CL limit on MM flux of $7.2\times 10^{-10}~\cmssr$ with an eight-loop SQUID~\cite{Huber:1990an} (cf.\ Fig.~\ref{fig:nova} in Section~\ref{sc:direct}).

A few years after the Stanford event observation, in August 1985, another unusual event was recorded this time at Imperial College London in a magnetometer with two superconducting loops~\cite{Caplin:1986kw}. It was observed during an operational period of total exposure-times-area corresponding to 400~times that of Cabrera's. Possible explanations, such as mechanical shock, residual magnetic flux or cosmic rays have been ruled out by the research team and it remains unexplained as of today. 

Another MM candidate had been seen earlier in 1973 at Berkeley by a team led by P.~B.~Price, in a balloon-borne Lexan\texttrademark\  emulsion stack~\cite{Price:1975zt}. The NTD and emulsion sheets   had recorded an unusual track corresponding to a HIP moving downward at near-relativistic speed. The track seemed consistent with a MM of charge 2\gd, velocity $\beta = 0.5$ and a mass of $\gtrsim 200~\gev$. However, comprehensive studies indicated that the track was probably due to the double fragmentation of a platinum nucleus~\cite{Alvarez:1975gm,Fowler:1975db}.

In 2006, the SLIM experiment (cf.\ Section~\ref{sc:slow-mms}) encountered a peculiar finding while analysing a CR-39\textsuperscript{\textregistered} foil, consisting of a sequence of  complicated ``tracks'' along a 20~cm line, dissimilar to usual ion tracks, such as the ones in Fig.~\ref{fig:etchpit}. After a meticulous study also involving neighbouring sheets and a series of etching campaigns, it appeared that the origin was an extremely rare manufacturing defect affecting $\mathrm{1~m^2}$ of CR39\textsuperscript{\textregistered}~\cite{Giorgini:2007fm}.

More recently, the Telescope Array Collaboration announced the detection of an ultra high energy cosmic ray of energy 244~EeV, which is intriguing as it exceeds the Greisin--Zatsepin--Kuzmin energy limit and it comes from the Local Void~\cite{TelescopeArray:2023sbd}. Highly energetic light MMs  have been proposed as being the origin of this event~\cite{Cho:2023krz,Frampton:2024shp}.  

Events compatible with the passage of a MM, as the aforementioned cases, showcase an interesting feature of MM searches. Due to their unique feature, the magnetic charge, their existence is expected to be inferred from the observation of singular events with practically no obvious background sources involved. This is in contrast to modern-era particle-physics experiments, where signals of physics beyond the SM (BSM) are expected to manifest themselves as statistically established excesses over accumulated background distributions, the latter being estimated frequently through data-driven methods. On the downside, these unique MM-like events are too rare to be accounted for or studied with real data. These different paradigms characterising MM searches are by no means a defect; it rather underlines the great caution with which such candidate events should be interpreted.   

\section{Searches in colliders}\label{sc:colliders}

Present and proposed future accelerators feature collisions at a centre-of-mass parton energy $(\sqrt{s})$ of ${\mathcal O}(1-10~\tev)$, thus it is virtually impossible to search for GUT monopoles in these machines. Nevertheless, searches have been carried out to detect direct or indirect signals of lighter MMs~\cite{Lee:2018pag,Alimena:2019zri,Knapen:2022afb,Mitsou:2021tti,Mitsou:2025qko,Mitsou2026}. Searches have been performed at hadron--hadron, electron--positron and lepton--hadron experiments, mostly directly using scintillation counters, gas chambers and NTDs, taking advantage of the MM high-ionisation power. Other analyses focus on exposed material for trapped MMs or peculiar magnetic-charge trajectories. In addition, virtual-monopole processes enhancing production rates of certain final states have also been considered as indirect probes of MMs.

The last few years, the monopole searches interest has shifted to the Large Hadron Collider (LHC)~\cite{Evans:2008zzb}, which is the largest and highest energy particle collider to-date~\cite{Bose:2022obr}. It was built at CERN between 1998 and 2008 in the existing Large Electron Positron (LEP) tunnel of 27~km in circumference beneath the France--Switzerland border near Geneva. The LHC primarily collides proton beams, with heavy ons being collided since its operation startup in 2010, too. Since then, it has collided protons to protons ($pp$) at record energies of $\sqrt{s}=7~\tev$ to 13.6~\tev during three running periods: Run~1 (2010--2012), Run~2 (2015--2018) and Run~3 (2022--2026). The heavy-ion programme is equally conspicuous, providing data in nuclei--nuclei, e.g.\ Pb--Pb, Xe--Xe, and proton--nuclei, e.g.\ $p$--Pb, $p$--O, collisions.

\subsection{Monopole production processes \label{sc:cross}}
 
\begin{figure}[ht]
     \centering
     \begin{subfigure}[b]{0.4\textwidth}
         \centering
         \includegraphics[width=0.7\textwidth]{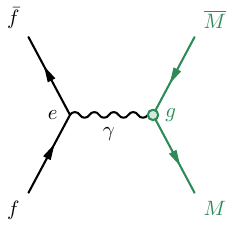}
         \caption{Drell--Yan production.}
         \label{fig:dy}
     \end{subfigure}
     \hspace{0.04\textwidth}
     \begin{subfigure}[b]{0.4\textwidth}
         \centering
         \includegraphics[width=0.7\textwidth]{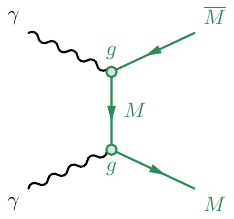}
         \caption{Photon-fusion production.}
         \label{fig:pf}
     \end{subfigure}
\caption{Monopole $M$ pair production diagrams in colliders. \subref{fig:dy} Drell--Yan production: The fermion $f$ can be  quarks, e.g.\ in a hadron collider, or leptons $(e^{\pm}, \mu^{\pm})$. \subref{fig:pf} Photon fusion: The photons may radiate off colliding fermions or nuclei, e.g.\ in ultraperipheral heavy-ion collisions.}
\label{fig:direct}
\end{figure}

Direct monopole pair production in colliders can proceed via two processes: a Drell--Yan-like (DY) process in photon $s$-channel intermediation (see Fig.~\ref{fig:dy}) and a photon-fusion (PF) $t$-channel diagram~\cite{Kurochkin:2006jr,Dougall:2007tt,Baines:2018ltl,Song:2021vpo} (see Fig.~\ref{fig:pf}). For both mechanisms, duality arguments justify an effective \bt-dependent magnetic charge in MM--matter scattering processes ---which consequently also characterises MM production--- with \bt defined as\footnote{This \bt parameter is totally unrelated to the monopole velocity $\beta$ discussed in Section~\ref{sc:cosmics}.}~\cite{Schwinger:1976fr,Milton:2006cp}
\be\label{defbeta}
\bt = \sqrt{1 - \frac{4M_{\mathrm m}^2}{s}},
\ee
where $M_{\mathrm m}$ is the monopole mass and $s$ is the Mandelstam variable. The $\bt$-independence, on the other hand, is justified within a framework of effective gauge field theory approach to MM--matter scattering~\cite{Alexandre:2026auj}. The issue is not settled yet.

An important cautionary remark is due here. In both cases, the MM pair couples to the photon via a coupling that depends on $\gd$ and hence it is ${\mathcal O}(10)$. This large MM--photon coupling invalidates \emph{in principle} the perturbative treatment of the cross-section calculation and hence the mass bounds  based on it are \emph{indicative}. On the contrary, the upper bounds placed on production cross sections are solid and can be safely relied upon~\cite{ParticleDataGroup:2024cfk}. This issue can be evaded (or resolved in its entirety) in a specific kinematical regime of the PF process, in the Schwinger mechanism, and through resummation techniques, as discussed in Sections~\ref{sc:kappa},~\ref{sc:schwinger}, and~\ref{sc:resum}, respectively.

\subsubsection{Magnetic moment in photon fusion}\label{sc:kappa}

A possibility to circumvent the point in question is the $\gamma\gamma$-fusion production for fermionic and vector boson MMs with a \bt-dependent coupling. A magnetic-moment term proportional to a phenomenological parameter $\kappa$ is added to the effective Lagrangians for spins \hf and~1, without affecting the validity of the DQC. The possibility to use the parameter $\kappa$ in conjunction with \bt to achieve a perturbative treatment of the MM--photon coupling was introduced in Ref.~\cite{Baines:2018ltl}. Indeed, by limiting the discussion to MMs with $\bt \ll 1$ , the perturbativity is guaranteed, however, at the expense of a vanishing cross section in DY production. It turns out that the \emph{photon-fusion} cross section remains finite \emph{and} the coupling is perturbative at the formal limits $\kappa\to\infty$ and $\bt\to 0$ when the following condition is met:
\be g\kappa\bt^2 < 1,
\label{eq:pert}\ee
provided a $\bt$-dependent magnetic charge is used~\cite{Milton:2006cp,Schwinger:1976fr}. An effective field theory for the MM--$\gamma$ interaction supporting this latter point of view is discussed in Section~\ref{sc:resum}.

\subsubsection{Schwinger mechanism}\label{sc:schwinger}

Pairs of electrically charged particles can be created in strong enough electric fields, a phenomenon known as the \emph{Schwinger mechanism}~\cite{Schwinger:1951nm}. By electromagnetic duality, a sufficiently strong magnetic field would similarly produce MM pairs~\cite{Affleck:1981ag,Drukier:1981fq}. The possibility of calculating its rate through semi-classical techniques~\cite{Gould:2017zwi,Gould:2017fve,Gould:2018efv,Gould:2019myj,Ho:2019ads,Gould:2021bre} overcomes the non-perturbativity of the MM--$\gamma$ coupling. Besides that, this approach evades the exponential suppression in the production cross section of \emph{composite} monopoles, which is a factor of~\cite{Drukier:1981fq}
\be e^{-4/\alpha} \simeq 10^{-238}, \quad \text{with }\: \alpha \simeq 1/137.
\label{eq:suppr}\ee
 
This mechanism becomes effective in the presence of strong electromagnetic fields, such as those present in heavy-ion collisions, and does not rely on perturbation theory, therefore, it overcomes the aforementioned  limitations~\cite{Gould:2017zwi,Gould:2017fve,Gould:2019myj,Ho:2019ads}.  As we shall see in Sections~\ref{sc:moedal-schwinger} and~\ref{sc:atlas-schwinger}, searches at LHC experiments have set constraints on the MM production using this mechanism. 

\subsubsection{Resummation}\label{sc:resum}

An effective $U(1)_\text{em} \times U(1)_\text{dual}$ gauge field theory for MM--$\gamma$ coupling has been developed, where $U(1)_\text{em}$ denotes the usual gauge group of Electromagnetism, while $U(1)_\text{dual}$ is a strongly coupled \emph{dual} gauge group~\cite{Alexandre:2019iub}. It uses a novel Dyson--Schwinger (DS) approach for resumming the MM--photon graphs of Fig.~\ref{fig:direct} in a gauge and Lorentz-invariant way, using formalism developed in Ref.~\cite{Terning:2018udc}. A similar model has been studied non-perturbatively using lattice techniques~\cite{Farakos:2024ggp}. This approach has its starting point in the two-potential formalism~\cite{Zwanziger:1970hk} for constructing a \emph{local} effective action for MMs, avoiding the use of Dirac strings.
 
Very recently, an extension of this one-loop DS resummation~\cite{Alexandre:2019iub} determined a non-trivial UV-fixed-point structure with the Lagrangian density assuming the form of a QED model in a specific gauge fixing~\cite{Alexandre:2026auj}. The self-consistent definition of the magnetic charge, obeying the DQC, left \emph{unchanged} the tree-level cross sections of the DY and PF production processes~\cite{Baines:2018ltl,Song:2021vpo} for spin-\hf MMs, thus \emph{justifying} their use in experimental MM searches at colliders and validating the extracted MM mass bounds, presented in Sections~\ref{sc:coll-direct}--\ref{sc:moedal}. The same holds for processes involving virtual MMs such as the light-by-light scattering, discussed in Section~\ref{sc:box}. In addition, such techniques support the existence of a velocity-effective magnetic charge. 

Consequently, the long-standing issue stemming from the large monopole--photon coupling is resolved. It is reminded, however, that higher-order resummation may result in modifications of the production cross sections and the resulting mass limits~\cite{Alexandre:2026auj}.  The same DS resummation approach has been applied in the past on BSM particles with very high electric charges~\cite{Alexandre:2023qjo,Alexandre:2024pbs,Musumeci:2025tuw}, as discussed in Section~\ref{sc:hecos}.

\subsection{Indirect searches}\label{sc:box}

\subsubsection{Via monopolium}\label{sc:mono}

\begin{figure}[ht]
     \centering
     \begin{subfigure}[b]{0.4\textwidth}
         \centering
         \includegraphics[width=0.7\textwidth]{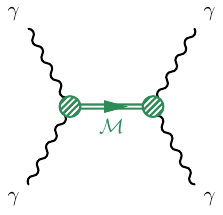}
         \caption{Monopolium production and decay.}
         \label{fig:mono}
     \end{subfigure}
     \hspace{0.04\textwidth}
     \begin{subfigure}[b]{0.4\textwidth}
         \centering
         \includegraphics[width=0.7\textwidth]{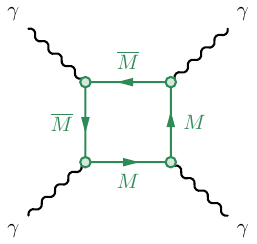}
         \caption{Virtual-monopole box diagram.}
         \label{fig:box}
     \end{subfigure}     
       \caption{Monopole-related diphoton production:  \protect\subref{fig:mono} monopolium $\mathcal{M}$ production via $\gamma\gamma$-fusion and subsequent decay into two photons, and \protect\subref{fig:box} light-by-light scattering through a virtual-monopole loop.}
\label{fig:indirect}
\end{figure}

Indirect searches for MMs may proceed by seeking the monopolium~\cite{Hill:1982iq,Dubrovich:2003mv,Vento:2007vy}, a bound state of a monopole and an antimonopole, previously discussed in Section~\ref{sc:monopolium}. This object can be produced via $\gamma\gamma$ fusion, as shown in Fig.~\ref{fig:mono}, in $e^+e^-$ annihilation~\cite{Epele:2007ic,Reis:2017rvb}, in high-energy proton--(anti)proton collisions~\cite{Ginzburg:1998vb,Epele:2008un,Epele:2012jn} and in heavy-ion collisions~\cite{daSilva:2023jxd}. Monopolium is a neutral state, hence it is difficult to detect directly at a collider detector, however, its decay into two photons would give a potentially detectable signal in LHC experiments~\cite{Epele:2012jn,Epele:2016wps}. Specific predictions for a $\sim 750$-\gev monopolium, motivated by signs of an excess in the diphoton-invariant-mass spectrum observed by ATLAS~\cite{ATLAS:2016gzy} and CMS~\cite{CMS:2016xbb} experiments\footnote{This enhancement was not consolidated by any of the two experiments when the analysed dataset was enlarged~\cite{ATLAS:2017ayi,CMS:2016kgr}, yet it prompted research teams to concretise the prediction power of their monopolium models.} benefitted posterior interpretations of collider searches results~\cite{Barrie:2016wxf,Epele:2016wps}. 

\begin{figure}[htb]
\sidecaption
\includegraphics[width=0.55\textwidth]{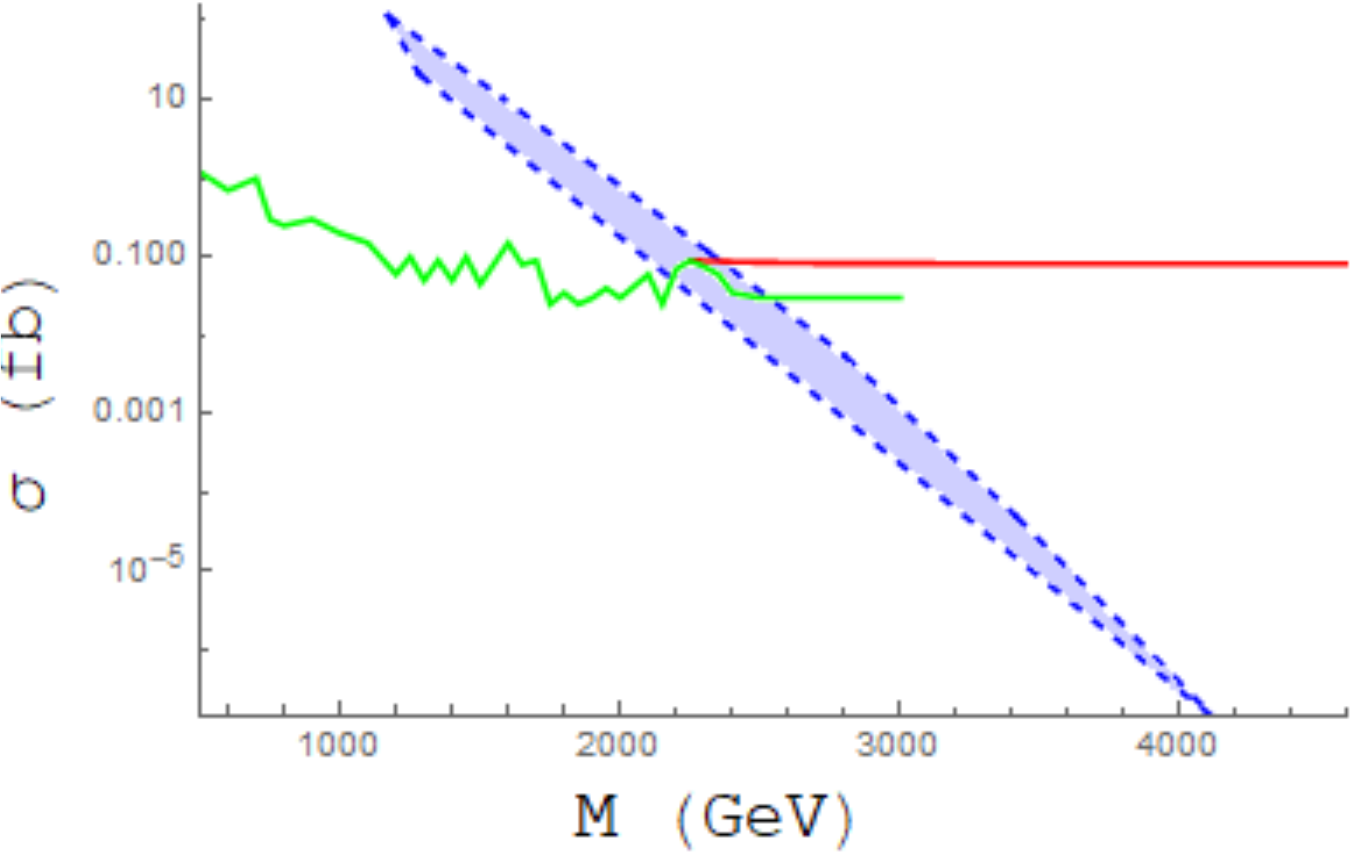}
\caption{Predicted diphoton cross section at 13~\tev LHC for the monopolium ground state and excited states, including multiphoton decay processes, versus the monopolium mass, for a MM mass of 4~\tev, juxtaposed on the 95\% CL upper limits from CMS (red)~\cite{CMS:2018dqv} and ATLAS (green)~\cite{ATLAS:2021uiz}. The blue region represents the range of predicted cross sections for various values of model parameters. From Ref.~\cite{Barrie:2021rqa}. \vspace{0.1cm}}
\label{fig:monopolium} 
\end{figure}

The monopolium has lower mass than the monopole--antimonopole pair and, depending on the binding energy, it may feature a much larger production cross section in LHC $pp$ collisions for the same MM mass. This is evident in the results of a study which considers a monopolium model based on the electromagnetic dual formulation of Zwanziger~\cite{Zwanziger:1968rs,Zwanziger:1970hk}. Diphoton production constraints set by ATLAS~\cite{ATLAS:2021uiz} and CMS~\cite{CMS:2018dqv} result in model-dependent exclusion of monopolium masses lower than $1\sim4~\tev$~\cite{Barrie:2021rqa}. As demonstrated in Fig.~\ref{fig:monopolium}, for a MM mass of 4~\tev, a (conservative) lower limit on the monopolium mass of 2.2~\tev is set. Moreover, monopole--antimonopole annihilation and a lightly bound monopolium may lead to multiphoton events (four and more photons in the final state), while for a strongly bound monopolium ---although diphoton events are dominant--- four- and six-photon event production is also sizeable~\cite{Fanchiotti:2017nkk,Barrie:2021rqa}. 

The existence of both monopole pairs and monopolia can be probed via the scattering of charged particles off them~\cite{Kazama:1976fm,Vento:2018sog,Vento:2019auh}. If they are produced in $pp$ collisions at the LHC, the beam particles interact with the effective magnetic dipole they represent and they are deflected in off-forward directions. Monopole--antimonopole pairs lead to a sizeable effect, sufficient for detection in ATLAS, CMS and LHCb~\cite{Vento:2019auh}. Other suggestions for monopolium detection is via excited monopolium states~\cite{Fanchiotti:2022xvx} and ionisation energy loss in a medium\cite{Fanchiotti:2023jmx}.

\subsubsection{Via virtual monopoles}\label{sc:virtual}

\begin{figure}[htb]
\sidecaption
\includegraphics[width=0.5\textwidth]{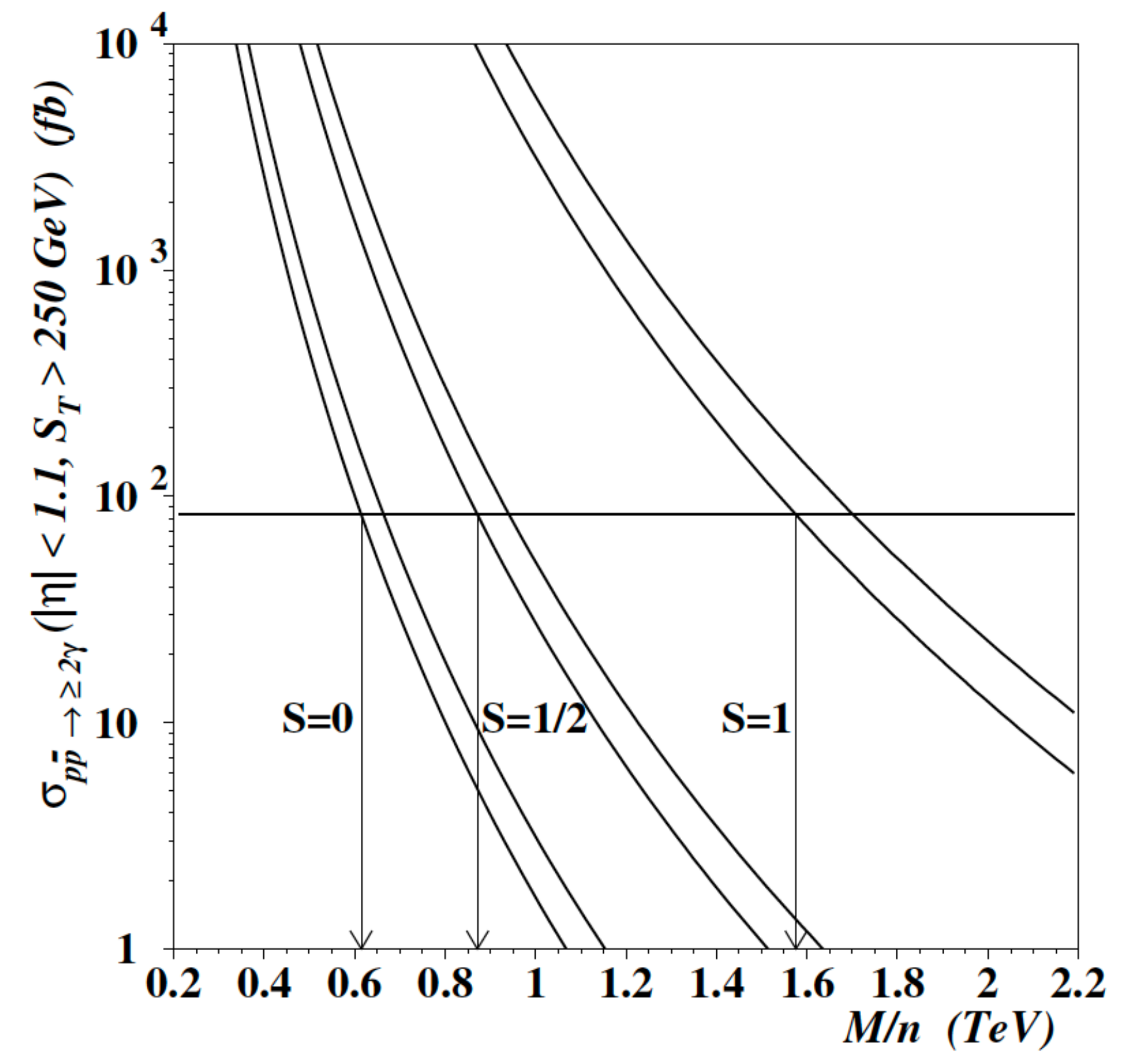}
\caption{Indirect constraints on MMs of charge $n\gd$ from D$\emptyset$~\cite{D0:1998ume}. The curved bands represent the theoretical cross sections with their uncertainties~\cite{Ginzburg:1998vb} for MM spin, $S = 0, \hf$, and~1. From the 95\% CL experimental upper limit on the cross section (horizontal line), the lower limits on MM mass divided by $n$ at each spin value are derived (arrows). From Ref.~\cite{D0:1998ume}. \vspace{0.2cm}}
\label{fig:dzero} 
\end{figure}

In addition, virtual monopoles have been suggested to mediate processes giving rise to multi-photon final states via the \emph{box diagram} shown in Fig.~\ref{fig:box}~\cite{DeRujula:1994nf,Ginzburg:1982fk}. Using dimension-8 effective field theory operators~\cite{Ginzburg:1998vb,Ginzburg:1999ej}, a diphoton-based search have been carried out by D$\emptyset$~\cite{D0:1998ume} in $p\bar{p}$ collisions at the Tevatron leading to the spin-dependent lower mass limits shown in Fig.~\ref{fig:dzero}. The L3 experiment at the LEP reported a lower mass limit of 510~\gev in a search of anomalous $Z\to\gamma\gamma\gamma$ events~\cite{L3:1994shn}. However, the uncertainties of the cross-section calculations used to derived these limits are difficult to estimate~\cite{Gamberg:1998xf,Milton:2006cp}. 

The observation of light-by-light scattering at the LHC by ATLAS~\cite{ATLAS:2017fur} and CMS~\cite{CMS:2018erd} in Pb--Pb ultraperipheral collisions (UPC) opened a new window for looking for MMs indirectly through loop contributions~\cite{dEnterria:2013zqi,Bruce:2018yzs}. A re-interpretation of the first ATLAS result led to a lower limit of 11~\tev on MM masses~\cite{Ellis:2017edi} in a Born--Infeld model~\cite{Born:1934gh}.  Analyses performed with LHC data involving two or more photons at the final stage, such as the measurement of $H \to \gamma\gamma$, may further constrain MMs~\cite{Epele:2012jn,Ellis:2016glu}. Diphoton measurements using forward-proton tagging to probe photon-initiated central exclusive production by CMS--TOTEM excluded anomalous contributions to this process~\cite{TOTEM:2023ewz}. Very recently, this result, expressed in terms of dimension-8 effective field theory (EFT) operators~\cite{Ginzburg:1998vb,Ginzburg:1999ej}, led to spin- and charge-dependent mass constraints, while a mass range of 10--60~\tev  of a MM of charge 2\gd in a Born--Infeld scenario was excluded~\cite{Mitsou:2026gvp,Musumeci:2025tuw}. Prospects for pursuing such analyses at future $e^+e^-$ colliders are promising~\cite{Ellis:2022uxv}. 

We stress here that constraints for MMs via precision measurements are subject to limitations related to EFT validity, perturbativity and unitarity conditions~\cite{Ginzburg:1999ej,Fichet:2013gsa,Fichet:2014uka} and may not replace direct searches. For instance, the EFT framework is only valid at energies \emph{below} the pair production threshold~\cite{Mitsou:2026gvp}, whereas direct searches take over \emph{above} this threshold. In this sense, indirect constraints act in a way complementary to explicit searches for MMs.

\subsection{Evolution of direct searches}\label{sc:coll-direct}

Several collider experiments utilised a variety of techniques to search for MMs in the past. In most of these analyses, results were expressed as upper limits on production cross sections versus the MM mass, under the ansatz that the kinematic properties of the MM-pair production were determined either by the DY process or by generic physics-process-agnostic distributions. Most analyses then proceeded to set mass limits assuming a (leading-order) DY total cross section, which is now validated by resummation schemes (cf.\ Section~\ref{sc:resum}). 

\subsubsection{Hadron colliders}\label{sc:tevatron}

The most recent direct searches for MMs at the Tevatron were carried out by the CDF~\cite{CDF:2005cvf} and the E882~\cite{Kalbfleisch:2000iz,Kalbfleisch:2003yt} experiments. The CDF collaboration used a time-of-flight system with a dedicated trigger requiring large light pulses in the scintillators and an offline selection requiring large \dedx tracks not curving in the plane perpendicular to the magnetic field. With $p\bar{p}$ collisions at $\sqrt{s}=1.96~\tev$, this analysis yielded a MM production cross-section limit of 0.2~pb for MM masses in the range 100--700~\gev and a mass limit of 360~\gev  for the DY process~\cite{CDF:2005cvf}.

The E882 experiment, also known as the ``Oklahoma experiment'', employed the induction technique (cf.\ Section~\ref{sc:squid}) to search for stopped MMs in  material, such as a $^9$Be beam pipe and other parts of the CDF and D$\emptyset$ detectors exposed to $\sim 175~\ipb$ of $p\bar{p}$ collisions~\cite{Kalbfleisch:2000iz,Kalbfleisch:2003yt}. Upper cross-section limits of 0.6, 0.2, 0.07 and 0.02~pb were obtained for magnetic charges of \gd, 2\gd, 3\gd and 6\gd, respectively, assuming a uniform MM-pair production. These bounds are translated into 265, 355, 410 and 375~\gev lower mass limits for a DY-like cross-section calculation~\cite{Kalbfleisch:2003yt}. An interesting study was performed by assuming three different angular distribution $(\mathrm{d}\sigma / \mathrm{d}\!\cos\theta)$ scenarios for the MM production: constant, $1+\cos^2\theta$ and $1-\cos^2\theta$. This variation leads to a $\pm 15~\gev$ spread in the obtained mass limits.  

Earlier searches~\cite{Price:1987py,Price:1990in,Bertani:1990tq} at $\sqrt{s}=1.8~\tev$ $p\bar{p}$ collisions at the Tevatron used NTDs to set an upper limit on the production of MMs with mass lower than 850~\gev at a cross section of 200~pb at 95\% CL~\cite{Bertani:1990tq}.  Lower-energy hadron--hadron experiments have employed a variety of search techniques including plastic track detectors at the CERN ISR~\cite{Hoffmann:1978mp} and $\rm Sp\bar{p}S$~\cite{Aubert:1982zi} and the extraction method to search for trapped monopoles (see also Section~\ref{sc:bound}) in proton beam dumps of 300~\gev~\cite{Carrigan:1973mw} and 400~\gev~\cite{Carrigan:1974un} at Fermilab and the CERN ISR~\cite{Carrigan:1977ku}.

\subsubsection{Electron--positron colliders}\label{sc:lep}

Regarding $e^+e^-$ colliders, the only LEP-2 search was made by OPAL based on 62.7~\ipb of data collected on average at $\sqrt{s} = 206.3~\gev$~\cite{OPAL:2007eyf}. By searching for two back-to-back particles with an anomalously high \dedx in the tracking chambers, an average upper limit of 0.05~pb was acquired on the MM-pair-production cross section in the mass range 45--102~\gev. Earlier at LEP-1, NTDs deployed around the interaction point, allowing probing high charges for masses up to \mbox{$\sim45~\gev$.} Specifically, the L6-MODAL experiment~\cite{Pinfold:1993mq} set limits on MMs with charges in the range 0.9--3.6\gd, while a previous search by the MODAL experiment was sensitive to magnetic charges as low as $0.1\gd$~\cite{Kinoshita:1992wd}. The deployment of NTDs around the beam interaction point was also used at $e^+e^-$ facilities of lower collision energies such as PEP~\cite{Kinoshita:1982mv,Fryberger:1983fa} at SLAC, PETRA~\cite{Musset:1983ii} at DESY and the TRISTAN ring~\cite{Kinoshita:1988cn,Kinoshita:1989cb} at KEK. 

\begin{figure}[htb]
\sidecaption
\includegraphics[width=0.55\textwidth]{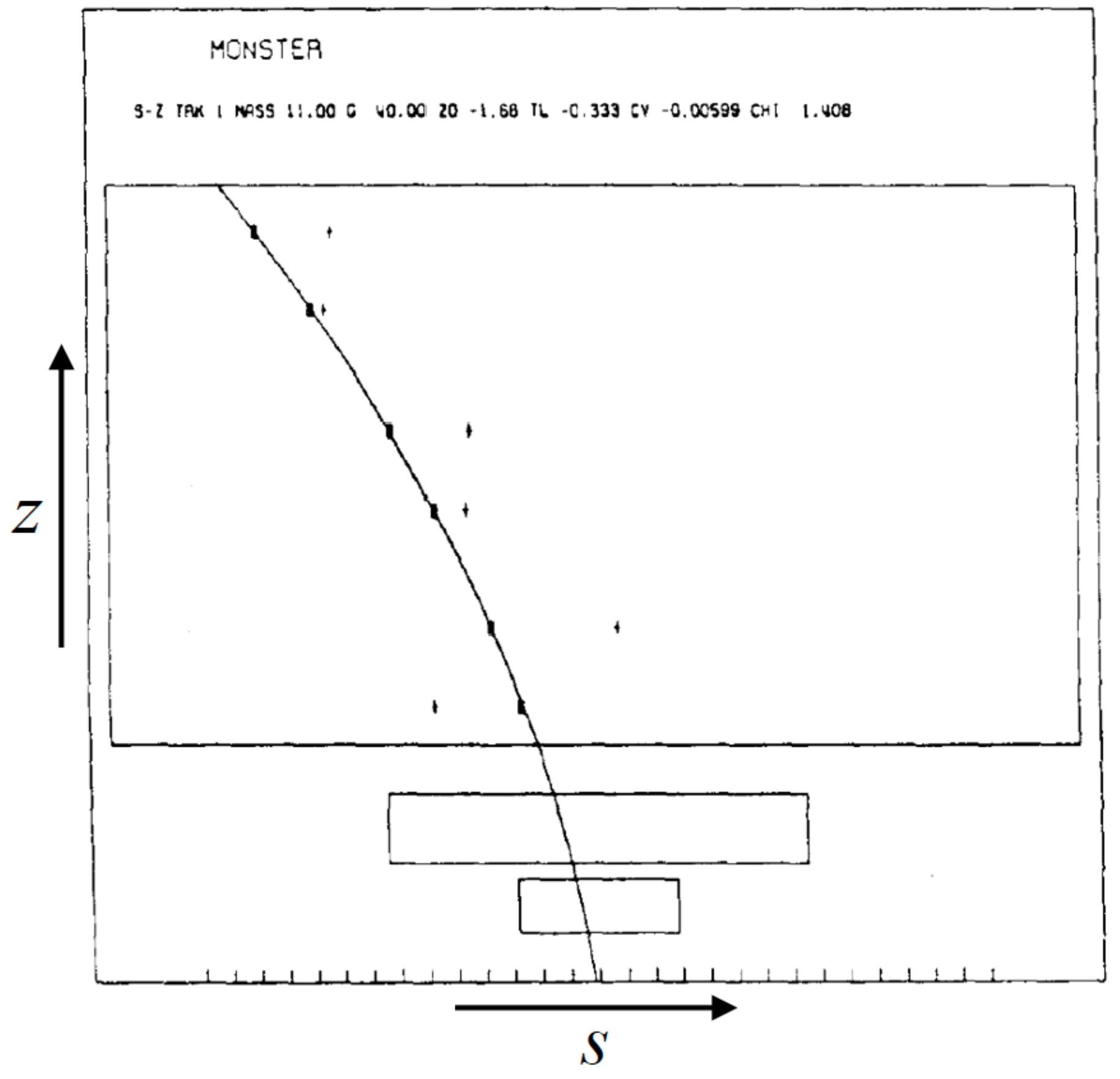}
\caption{Typical simulated MM event with mass 11~\gev and charge $g = 40e$ from TASSO. The $s-z$ view of one of the tracks is shown, where $z$ is the beam axis and $s$ is the orthogonal-to-$z$ axis in the plane defined by the two trajectories. The track is fitted with a parabola. Adapted from Ref.~\cite{TASSO:1988txx}. \vspace{0.3cm}}
\label{fig:tasso} 
\end{figure}

A different MM detection technique, which relies of non-helical trajectories, has been applied at $e^+e^-$ colliders in the CLEO~\cite{CLEO:1986lye} detector at Cornell and the TASSO~\cite{TASSO:1988txx} detector at PETRA. The track of a simulated MM in the $s-z$ plane, fitted by a parabolic function, is shown in Fig.~\ref{fig:tasso} for the TASSO analysis. Cross section upper limits on the process $e^+e^- \to M \overline{M}$ for MMs with chargees $\leq1\gd$ and masses up to 16~\gev were set~\cite{TASSO:1988txx}. 

It is stressed that the non-helical-trajectory method, unlike the majority of other techniques, requires no additional detection apparatus for a collider experiment. It does necessitate, however, the development of dedicated pattern recognition, track reconstruction algorithms, and possibly a specialised trigger. The track fitting precision is expected to be inferior than for electric charges and the (transverse) momentum measurement will be deteriorated if existent. This is due to experiments being typically designed and optimised for electrically charged particles, e.g.\ in a tracker operating in a solenoid magnetic field, silicon microstrips and drift tubes are oriented parallel to the beam axis.

\subsubsection{Lepton--hadron colliders}\label{sc:hera}

\begin{figure}[htb]
\sidecaption
\includegraphics[width=0.5\textwidth]{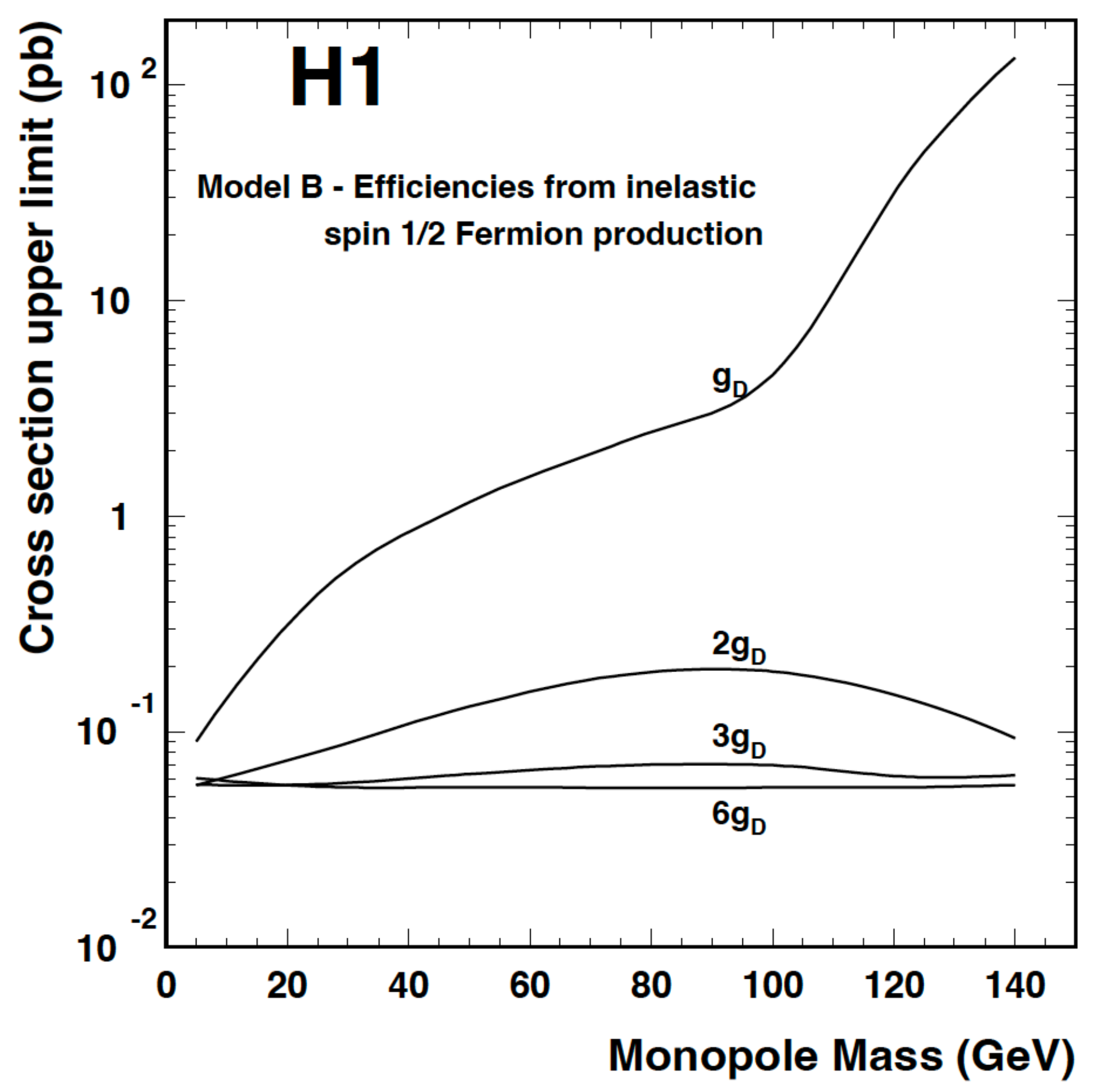}
\caption{Upper limits on the cross section for \mbox{spin-\hf} MM pair production in $e^+p$ collisions from H1, as a function of MM mass for magnetic charges of 1\gd,   2\gd, 3\gd, and 6\gd or more. From Ref.~\cite{H1:2004zsc}. \vspace{0.1cm}}
\label{fig:h1} 
\end{figure}

Up to now the only search for MM production in lepton--hadron scattering used the induction method on the $^{27}$Al beam pipe used by the H1 experiment at HERA exposed to $e^+p$ collisions at $\sqrt{s} = 300~\gev$~\cite{H1:2004zsc}. Parts of the beam pipe were scanned by a SQUID magnetometer with a sensitivity as low as 0.1\gd and no MM candidates were found. Pair production of spin~\hf MM was considered through the inelastic process $e^+ p \to e^+ M \overline{M} X$, where $X$ is any state, proceeding via a $\gamma\gamma$-fusion interaction with one photon radiating from the electron and the other radiating from a quark in the proton. With an integrated luminosity of 62~\ipb, upper limits on the MM pair production cross section were set  for magnetic charges in the range 1--6\gd and for masses up to $\sim 140~\gev$, as shown in Fig.~\ref{fig:h1}. Similar results were obtained for scalar MMs produced in the process $e^+ p \to e^+ M \overline{M} p$. The employment of the induction method, one of the few with which charges higher than 1\gd can be probed relatively effortlessly, has been revived at the LHC, as discussed thoroughly in Section~\ref{sc:lightsearch}.

\subsubsection{Towards the Large Hadron Collider}\label{sc:lhc}

Nowadays, searches for monopoles produced at the highest available energies in hadron--hadron collisions are being carried out in $pp$ and UPC collisions at the LHC~\cite{Evans:2008zzb} by the MoEDAL~\cite{MoEDAL:2009jwa} and ATLAS~\cite{ATLAS:2008xda} experiments utilising entirely different detection techniques, as elaborated in Sections~\ref{sc:moedal} and~\ref{sc:atlas}, respectively. 

At CMS~\cite{CMS:2008xjf}, a special track-combining algorithm to reconstruct the MM track and extrapolate it to the nearest crystal of the electromagnetic calorimeter has been developed. MM candidates are identified by high ionisation in the tracker and narrow energy deposition in the calorimeter~\cite{HassanElSawy:2019oxk} or by large missing transverse momentum~\cite{MenezesdeOliveira:2025edl}, providing a sensitivity less competitive than ATLAS and MoEDAL.  

\subsection{The MoEDAL experiment}\label{sc:moedal}

The Monopole and Exotics Detector at the LHC (MoEDAL)~\cite{MoEDAL:2009jwa} has been designed to search for manifestations of BSM physics through HIPs in a manner complementary to the main LHC experiments~\cite{DeRoeck:2011aa}. Its main motivation is to pursue the quest for MMs at LHC energies, as well as for any massive, long-lived, slow-moving exotic particle~\cite{Fairbairn:2006gg} with single or multiple electric charge~\cite{MoEDAL:2014ttp}.

\subsubsection{The MoEDAL detector}\label{sc:moedal-det}

The MoEDAL detector~\cite{MoEDAL:2009jwa,MoEDAL:2014ttp} is deployed around the intersection region at Point~8 (IP8) of the LHC in the LHCb experiment~\cite{LHCb:2008vvz} Vertex Locator cavern. It is a unique and largely passive LHC detector comprising three different detection technologies. 

\begin{description}
\item[\textbf{Nuclear track detectors}] The main subdetector system is made of a large array of CR-39\textsuperscript{\textregistered},  Makrofol\textsuperscript{\textregistered} and Lexan\texttrademark\ NTD stacks surrounding the intersection area. The analysis procedure outlined in Section~\ref{sc:ntd} is followed for each plastic sheet and then each one is scanned looking for aligned etch pits~\cite{MoEDAL:2021mpi}.

Another type of (relatively high-threshold) NTD is the Very High Charge Catcher (threshold $z/\beta\sim50$), consisting of  two light stacks of Makrofol\textsuperscript{\textregistered}, deployed in the LHCb (forward) side  between RICH1 and the Trigger Tracker. 

\item[\textbf{Magnetic trappers}] A unique feature, for an LHC experiment,  is the use of magnetic monopole trappers (MMTs) to capture magnetically charged HIPs. The $^{27}$Al absorbers of MMTs are subject to an analysis looking for MMs or dyons at a remote SQUID magnetometer facility~\cite{DeRoeck:2012wua}, applying the induction technique described in Section~\ref{sc:squid}. 

\item[\textbf{TimePix radiation monitors}] The only active subdetector is a TimePix pixel array, which forms a real-time radiation monitoring system of beam-related backgrounds, such as spallation products. The time-over-threshold mode allows a 3D mapping of the charge spreading in the silicon-sensor volume, thus differentiating between various particles species from mixed radiation fields and measuring their energy deposition~\cite{MOEDAL:2021mrd,Bergmann:2026uxe,MoEDAL-timepix:2026}.

\end{description}

MoEDAL has been designed specifically to overcome limitations on the detection of heavy HIPs present in conventional high-energy experiments. It does not rely on readout boards, which would saturate as the high-energy loss exceeds the dynamic range specified for typical readout chips, leading to saturation of the detector electronics. Signal from heavy ---therefore slow-moving--- particles may arrive in detectors in bunch crossings subsequent to that of hard scattering, necessitating dedicated triggers and reconstruction algorithms,\footnote{Notwithstanding, the LHC experimental community has pursued a wide programme of specialised data-processing algorithms and tools to enhance ATLAS, CMS and LHCb sensitivity to long-lived particles~\cite{ATLAS:2013bsk,Alimena:2019zri,Alimena:2021mdu,Alimena:2025kjv}.} whereas MoEDAL is time-agnostic. Finally, MoEDAL operates triggerless, without requiring the development of special HIP triggers. 

\subsubsection{Monopoles and dyons in DY and PF processes}\label{sc:lightsearch}

\begin{figure}[htb]
\sidecaption
\includegraphics[width=0.55\textwidth]{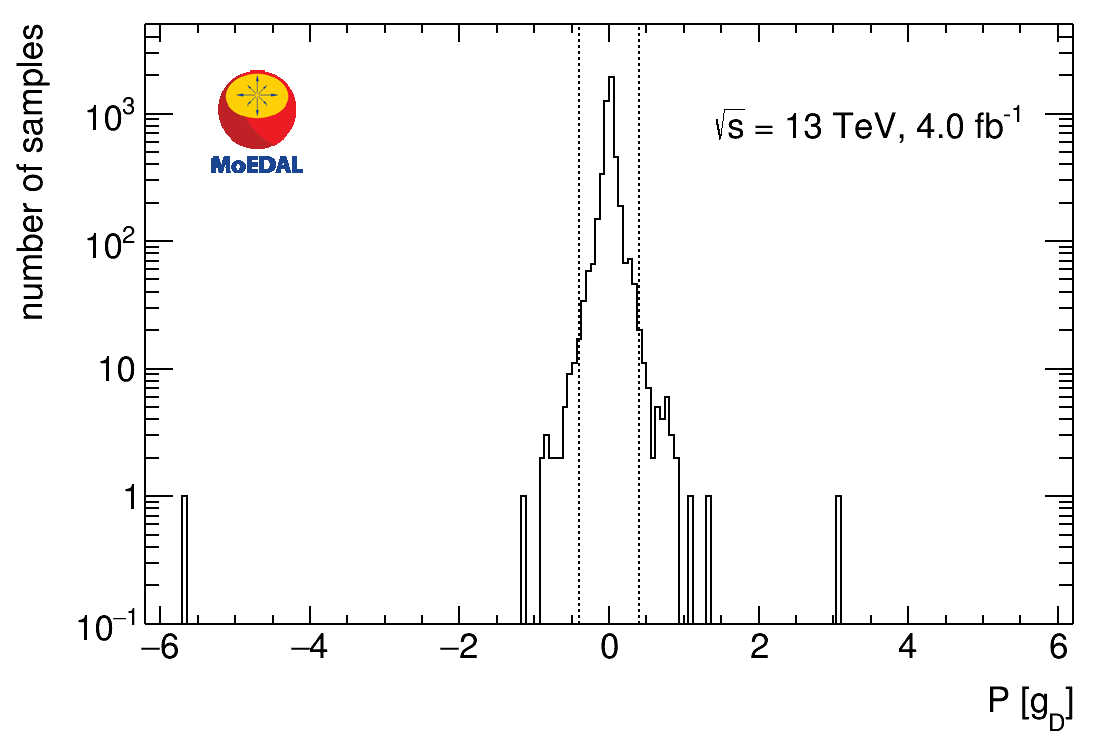}
\caption{MoEDAL SQUID analysis results~\cite{MoEDAL:2019ort}: Magnetic pole strength (in units of Dirac charge, \gd) measured through the induced persistent-current change in the 2400 $^{27}$Al samples of the MMTs exposed to 13~TeV $pp$ collisions in \mbox{2015--2017} with every sample scanned twice. From Ref.~\cite{MoEDAL:2019ort}. \vspace{0.2cm}}
\label{fig:squid-results} 
\end{figure}

The first MoEDAL physics results were based on the scanning of  MMTs, exposed to LHC Run-1 data at $\sqrt{s}=8~\tev$~\cite{MoEDAL:2016jlb} and to Run-2 13-\tev $pp$ collisions~\cite{MoEDAL:2016lxh,MoEDAL:2017vhz,MoEDAL:2019ort,MoEDAL:2020pyb,MoEDAL:2021mpi,MoEDAL:2023ost}. As shown in Fig.~\ref{fig:squid-results}, the persistent-current measurements were compatible with the non-observation of MMs and, therefore, upper limits on MM production cross sections were established. Subsequently, etching and scanning of NTDs exposed to 8-\tev~\cite{MoEDAL:2021mpi} and~13-\tev~\cite{MoEDAL:2023ost} LHC $pp$ collisions did not yield any HIP candidate either. As a result, analyses considering both NTDs and MMTs set even more stringent on MM mass of up to 4.2~\tev for magnetic charges as high as 10\gd~\cite{MoEDAL:2023ost}. MMs with very low charges tend to punch through MMTs, whereas higher charges may be absorbed by upstream material, thus limiting sensitivity. 

MoEDAL performed the first dedicated dyon search~\cite{MoEDAL:2020pyb} in a collider experiment by means of MMT scanning, which is sensitive to the \emph{magnetic} charge of the monopole. Mass limits in the range 750--1910~\gev were set using a benchmark DY production model for dyons with magnetic charge up to 5\gd, for electric charge from $1e$ to $200e$, and for spins~0,~\hf and~1. A schematic overview of the various dyon mass limits is given in Ref.~\cite{Mitsou:2022vka}. Previous results from analyses seeking HIPs in collider experiments could, in principle, be interpreted in terms of dyons using as a handle either the electric or the magnetic charge. In such cases, however, caution has to be exercised due to the non-trivial behaviour of dyons in the presence of magnetic field. 

Overall, MoEDAL pioneered monopole searches in several ways. It considered for the first time \mbox{spin-1} MMs~\cite{MoEDAL:2017vhz} and the PF production mechanism~\cite{Baines:2018ltl,MoEDAL:2017vhz,MoEDAL:2019ort,MoEDAL:2020pyb}. The dependency of the MM--photon coupling on the \bt parameter, defined in \eqref{defbeta} in Section~\ref{sc:cross}, has been  taken into account in several  MoEDAL results interpretations. 

\subsubsection{Schwinger-produced MMs}\label{sc:moedal-schwinger}

As discussed in Section~\ref{sc:schwinger}, the thermal Schwinger creation of MMs in strong magnetic fields is independent of the non-perturbative MM--$\gamma$ coupling and is valid both for structureless and for extended MMs. Heavy-ion collisions at the LHC produce the strongest known magnetic fields in the current Universe; more than four orders of magnitude greater than the strongest known astrophysical magnetic fields,  present on the surface of magnetars~\cite{Gould:2017zwi}.

Two approximations in the calculation of the MM production cross section have been considered in searches~\cite{MoEDAL:2021vix}.
\begin{description}
\item[\textbf{Free-particle approximation (FPA)}] The spacetime dependence of the electromagnetic field of the heavy ions is treated exactly, but MM self-interactions are neglected~\cite{Gould:2021bre}. 
\\
\item[\textbf{Locally constant field approximation (LCFA)}] The spacetime dependence of the electromagnetic field is neglected, but MM self-interactions are treated exactly~\cite{Gould:2019myj}. 
\end{description} 

\begin{figure}[ht]
\centering
\includegraphics[width=0.65\textwidth]{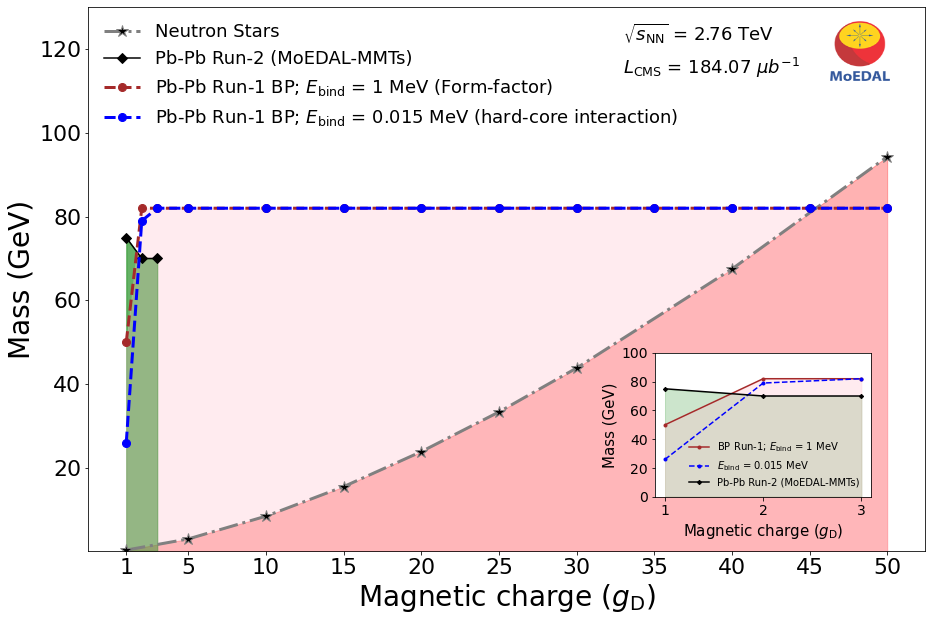}
\caption{Exclusion region for a MM search performed by MoEDAL via the Schwinger effect in the CMS beam pipe exposed to Pb--Pb collisions during LHC Run~1~\cite{MoEDAL:2024wbc}. The green-shaded region shows the MoEDAL MMT Run-2 limits~\cite{MoEDAL:2021vix}. The inset zooms in on the low-charge region. The limit from indirect searches for MMs produced by neutron stars~\cite{Gould:2021bre} is also shown. From Ref.~\cite{MoEDAL:2024wbc}. \vspace{0.1cm}}
\label{fig:beampipe} 
\end{figure}

The first search for such production was conducted with MMTs  exposed to 0.235~nb$^{-1}$ of 5.02~\tev/nucleon Pb--Pb UPC at IP8. They were later analysed with a SQUID, without finding any candidate~\cite{MoEDAL:2021vix}.  MMs with Dirac charges $1\gd \leq g \leq 3\gd$ and masses up to 75~\gev were excluded, as seen in the inset of Fig.~\ref{fig:beampipe}. This analysis provided the first lower mass limits for \emph{finite-size} MMs from a collider search, since the Schwinger mechanism is not subject to the exponential suppression that their cross section suffers in DY and PF processes, as discussed in Section~\ref{sc:schwinger}. 

In 2019, the CMS collaboration officially donated its Run-1 beam pipe to MoEDAL with the aim to search for captured MMs. Following the lead of similar analyses carried out at hadron colliders and at HERA (cf.\ Section~\ref{sc:coll-direct}), the $^9$Be beam pipe, which had been exposed to \mbox{2.76-\tev} Pb--Pb collisions, was sliced in pieces and scanned through a SQUID without finding any MM~\cite{MoEDAL:2024wbc}. The use of a trapping volume, e.g.\ beam pipe, very close to the collision point enhanced sensitivity to high charges, otherwise hindered by absorption in passive materials before reaching the detectors~\cite{DeRoeck:2011aa}. The result was interpreted via the Schwinger mechanism allowing setting the first reliable, world-leading mass limits on MMs with very high magnetic charge. In particular, the established limits are the strongest available in the range 2--45\gd, excluding MMs with masses of up to 80~\gev, as presented in Fig.~\ref{fig:beampipe}.  Recently MoEDAL has embarked in a re-interpretation of the non-observation of MMs in the CMS beam pipe considering the PF production in UPC heavy-ion collisions~\cite{photon-fusion-hi}.

\subsection{ATLAS searches}\label{sc:atlas}

ATLAS (A Toroidal LHC ApparatuS)~\cite{ATLAS:2008xda} is the largest, general-purpose particle detector experiment at the LHC, designed to measure the broadest possible range of SM properties, including the Higgs boson, and search for BSM signals~\cite{ATLAS:1999vwa}.

The ATLAS experiment consists of an inner tracking detector~\cite{ATLAS:2010ylv} surrounded by a thin superconducting solenoid, electromagnetic and hadronic calorimeters, and a muon spectrometer. The inner detector comprises  silicon pixel~\cite{Aad:2008zz}, silicon microstrip~\cite{Mitsou:2011gn,ATLAS:2021zxb}, and transition radiation tracking (TRT)~\cite{Mitsou:2003rp,ATLAS:2017jwu} detectors. Lead/liquid-argon (LAr) sampling calorimeters provide electromagnetic (EM) energy measurements~\cite{Abat:2009zz} with high granularity, while a steel/scintillator-tile hadronic calorimeter in the central region and LAr calorimeters provide electromagnetic and hadronic energy measurements~\cite{ATLASSecretariat:2010mep}. The muon spectrometer surrounds the calorimeters and is based on three large superconducting air-core toroidal magnets with eight coils each, and includes a system of precision tracking chambers and fast detectors for triggering~\cite{ATLAS:1997ad}.

\subsubsection{Based on TRT and electromagnetic calorimeter}\label{sc:atlas-dy-pf}

In ATLAS, searches for MMs have been performed at 7~\tev~\cite{ATLAS:2012bda}, at 8~\tev~\cite{ATLAS:2015tyu} and at 13~\tev~\cite{ATLAS:2019wkg,ATLAS:2023esy} data using the excellent TRT sensitivity to high-ionisation signals.\footnote{The ATLAS TRT transition-radiation feature aimed to provide particle identification capabilities~\cite{Akesson:2001su,ATLASTRT:2004lrl}. The accompanying two-threshold design of its readout electronics~\cite{Abat:2008zzc} produced a serendipitous sensitivity to highly ionising particles.} The latest 13~\tev analysis used 138~\ifb of data collected between 2015 and 2018 relied on a dedicated trigger for HIPs~\cite{ATLAS:2023esy}. The discriminating variables used in this search were the energy dispersion in the EM calorimeter~\cite{ATLASSecretariat:2010mep,Abat:2011zz}, $w$, and the fraction of TRT hits passing a predefined high threshold (HT), $f_\text{HT}$~\cite{ATLASTRT:2008rjy,Abat:2008zzc}. The energy dispersion measures the fraction of the cluster energy contained in the most energetic cells of a cluster in each of the layers of the electromagnetic calorimeter, in order to identify the characteristic ``pencil-shape'' energy deposit of the signal. Figure~\ref{fig:atlas-hips} shows the two-dimensional distribution of $w$ and $f_\text{HT}$ for data and a signal with 1\gd~\cite{ATLAS:2023esy}. It is evident that the variables are well discriminating and data in regions B, C and D can be used to predict the number of background events in the signal region A. 

\begin{figure}[htb]
\sidecaption
\hspace*{-0.02\textwidth}
\includegraphics[width=0.6\textwidth]{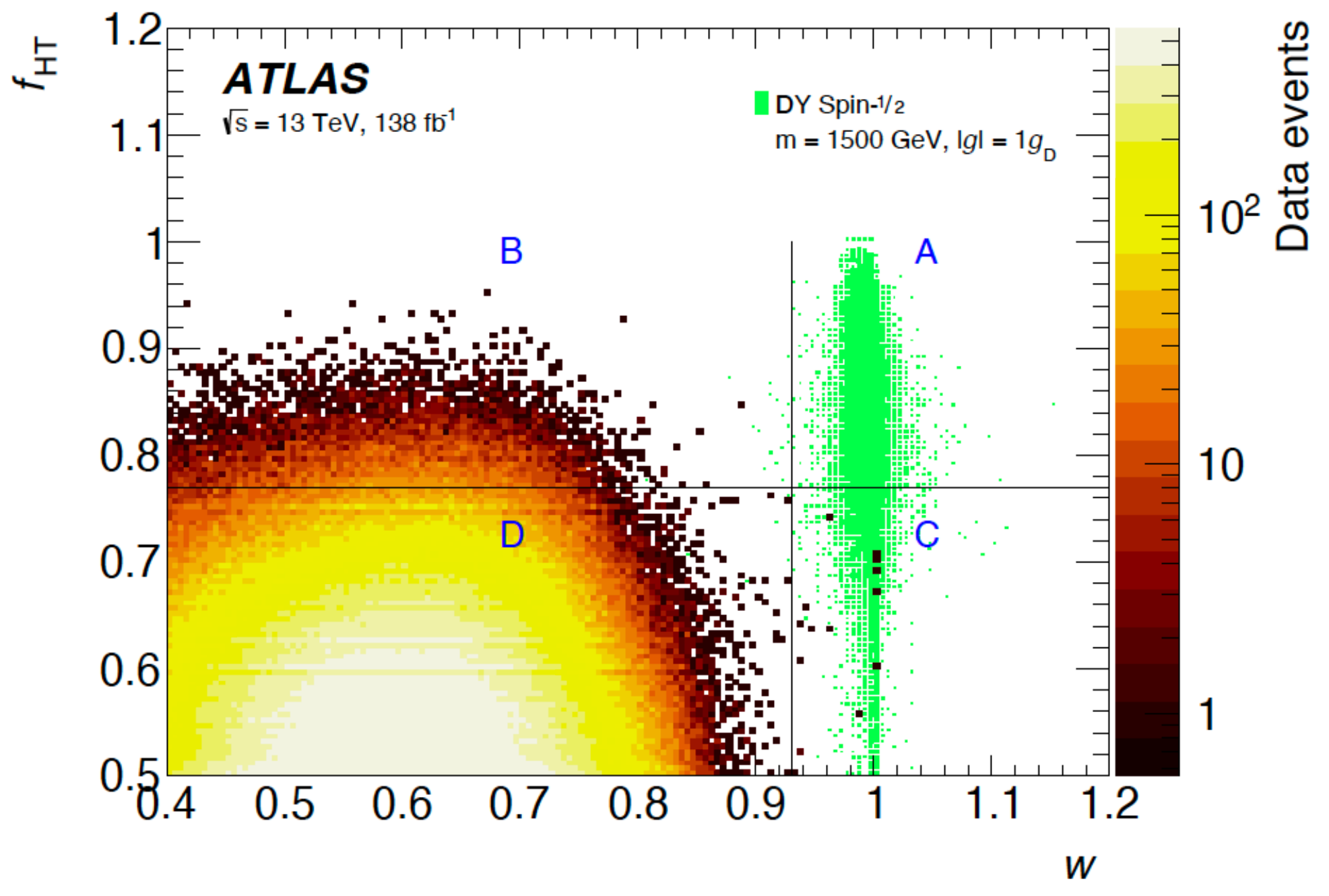}
  \caption{ATLAS HIP search~\cite{ATLAS:2023esy}. 2D distribution of discriminators $f_\text{HT}$ and $w$ for data and a signal model (green). The signal (A) and control (B, C and D) regions are shown.  From Ref.~\cite{ATLAS:2023esy}. \vspace{0.3cm} }
\label{fig:atlas-hips}
\end{figure}

No excess of data events was observed and this search result was interpreted assuming the DY and PF production processes, the latter considered for the first time in ATLAS. The analysis was sensitive to magnetic charges of $1\gd\leq|g|\leq2\gd$ and set limits for spin-0 and spin-\hf monopoles. The search excluded MMs up to masses of 3.7~\tev depending on the spin and production mechanism~\cite{ATLAS:2023esy}. These limits are the most stringent bounds placed by an LHC experiment on magnetic charges of $1\gd\leq|g|\leq2\gd$ to-date. 

The lower limits on MM mass obtained by MoEDAL and ATLAS are summarised in Fig.~\ref{fig:atlas-moedal-dy-pf}. Given that the production cross section for PF is much higher than the DY at the LHC energies~\cite{Baines:2018ltl,Song:2021vpo}, we observe that the PF mass limits dominate over the DY ones for the mass range of interest. 
The ATLAS bounds in low charges are better that the ones obtained by MoEDAL partly due to the higher luminosity delivered to ATLAS than MoEDAL.\footnote{A factor of $\sim 50$ less luminosity was delivered to MoEDAL than ATLAS during LHC Run~2 due to the operation requirements of the LHCb experiment. This factor has been reduced significantly during Run~3.} On the other hand, higher charges are difficult to be probed in ATLAS due to limitations imposed by the trigger and mainly to the saturation of the readout electronics used in such searches. MoEDAL remains the sole detector sensitive to high magnetic charges. The complementarity between the two approaches, a dedicated HIP-optimised detector (MoEDAL) and a powerful wide-spectrum experiment (ATLAS) is evident. 

\begin{figure}[htb]
\centering
  \includegraphics[width=0.8\linewidth]{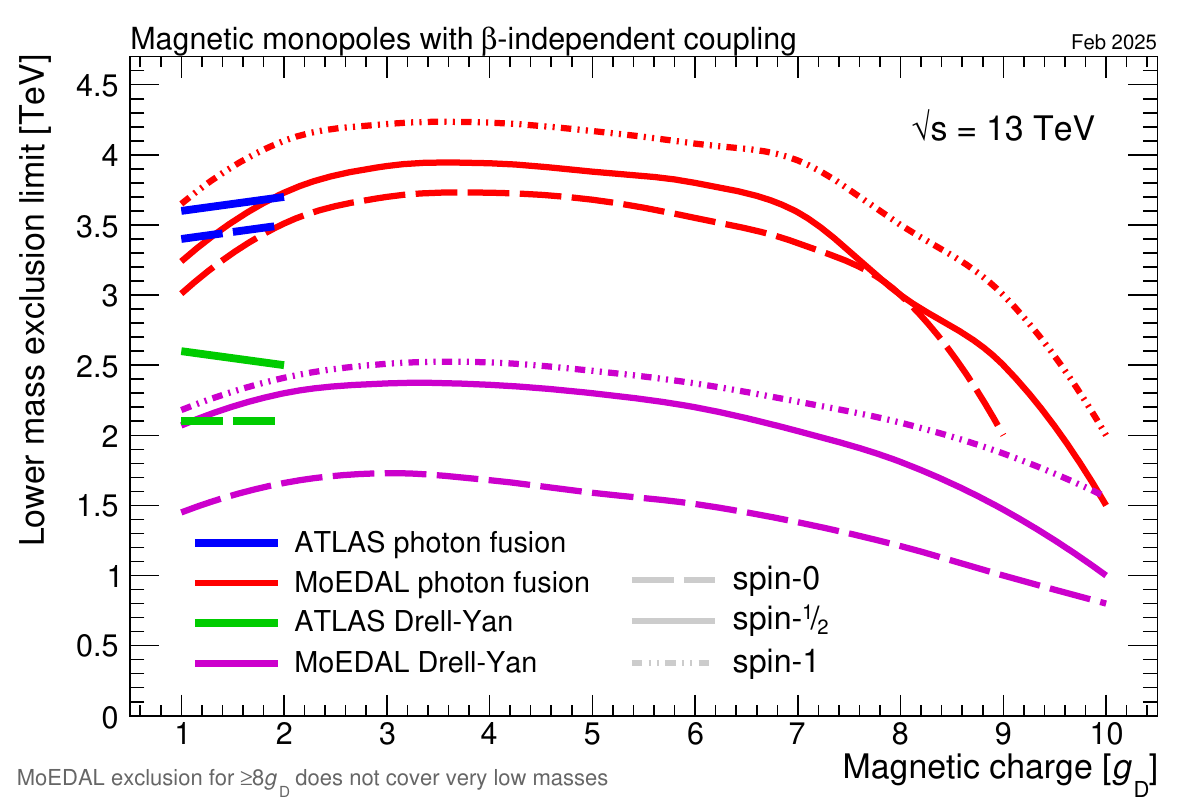}
  \caption{MM mass limits obtained by MoEDAL~\cite{MoEDAL:2023ost} and ATLAS~\cite{ATLAS:2023esy}  with $pp$ collisions at 13~\tev as a function of the magnetic charge. Drell--Yan and photon-fusion production with a $\beta$-independent coupling for monopoles of spin~0,~\hf and~1 is assumed. From Ref.~\cite{Mitsou:2025ufr}.} \label{fig:atlas-moedal-dy-pf}  
\end{figure}

\subsubsection{Based on Pixel detector}\label{sc:atlas-schwinger}

ATLAS has performed a search for monopole-pair production in UPC Pb--Pb collisions based on 5.36~\tev data recorded in 2023~\cite{ATLAS:2024nzp}. Due to their high ionisation and unique trajectories in a solenoidal magnetic field, MMs are expected to leave numerous clusters in the innermost  Pixel detector without associated reconstructed charged-particle tracks or calorimeter activity. The selection criteria exploit the straight track that a magnetic charge leaves on the plane perpendicular to the beam due to the solenoidal magnetic field unlike the curved track of any electric charge traversing the tracker. This leads to events with multiple pixel clusters~\cite{Aad:2008zz} and no or few reconstructed charged-particle tracks~\cite{Abat:2009zz,ATLAS:2010ylv,ATLAS:2014lvy}. Allowing for at most one calorimeter-cell energy deposit~\cite{ATLAS:2016krp} and requiring high transverse thrust further suppresses beam-induced background.

Monopoles of magnetic charge of 1\gd  in the mass range of 20--150~\gev have been excluded, as depicted in Fig.~\ref{fig:atlas-schwinger}. Both cross-section approximations, FPA and LCFA (cf.\ Section~\ref{sc:moedal-schwinger}), have been considered to extract lower mass limits; a conservative mass bound of 80~\gev has been set assuming FPA production rate. The ATLAS bounds are stronger\footnote{However, the collision energy in the MoEDAL analysis is slightly lower than that of ATLAS.} that those from MoEDAL, however, they are limited to MMs with charge 1\gd.

\begin{figure}[htb]
\sidecaption
\includegraphics[width=0.6\textwidth]{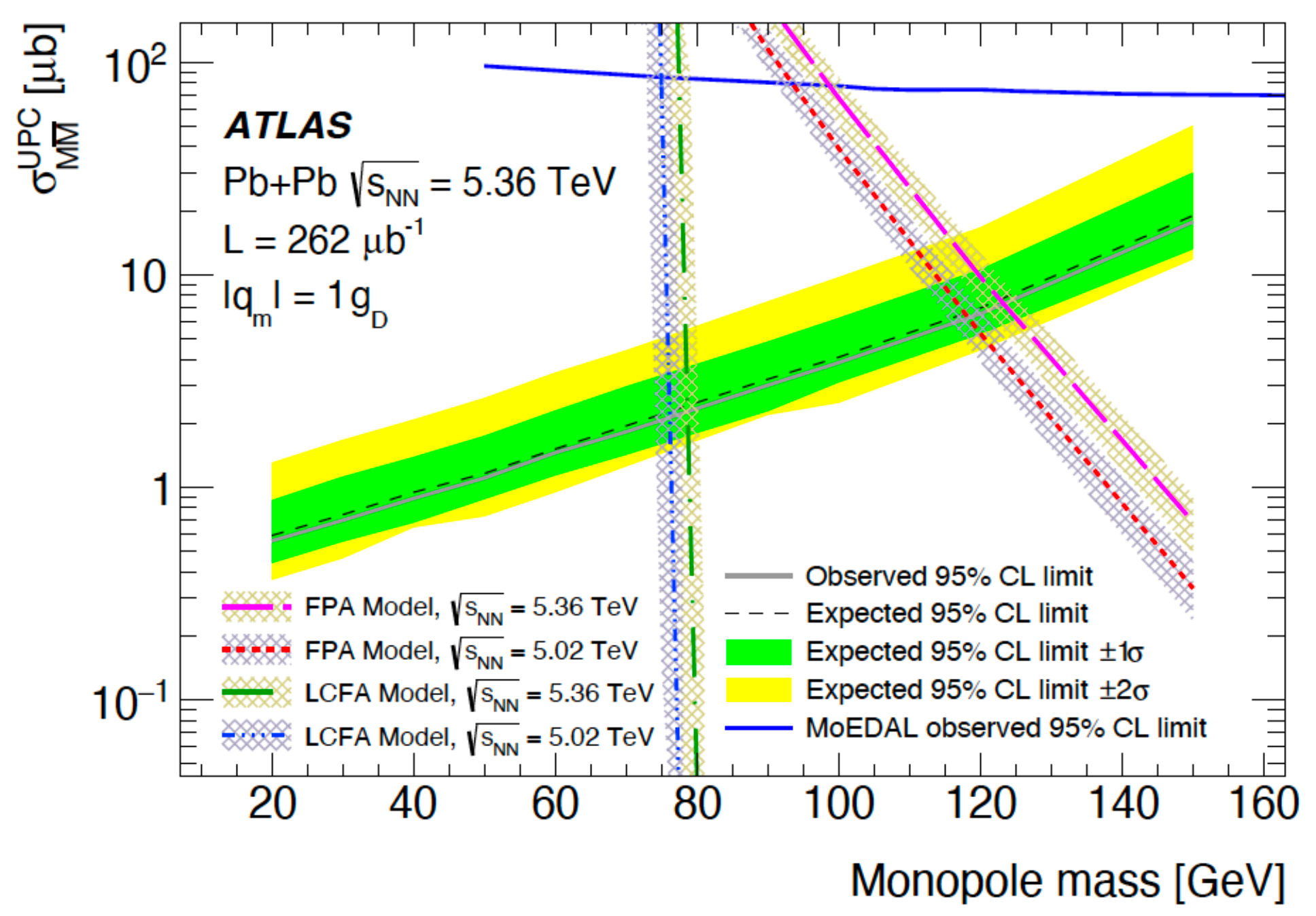}
\caption{ATLAS upper limits on 1-\gd MM-pair production cross-section in Pb--Pb UPC at $\sqrt{s_\text{NN}}=5.36~\tev$~\cite{ATLAS:2024nzp}. The grey solid line (black dashed line) represents observed (expected) limits, whereas the green (yellow) shaded bands are $\pm1\sigma$ ($\pm2\sigma$) intervals. The observed limits by MoEDAL at $\sqrt{s_\text{NN}}=5.02~\tev$  (blue line) and the FPA and LCFA model predictions (dashed/dotted lines) for both $\sqrt{s_\text{NN}}$ values are also shown. From Ref.~\cite{ATLAS:2024nzp}. \vspace{0.1cm}}
\label{fig:atlas-schwinger} 
\end{figure}

\section{Magnetic monopole searches constrain electric charges}\label{sc:hecos}

Several techniques using the high ionisation as a means to detect MMs, such as the ones presented in Section~\ref{sc:ionis}, may also be sensitive to massive, long-lived, slow-moving exotic particles with single or multiple electric charge~\cite{Fairbairn:2006gg,Mitsou:2024tyq,Maselek:2025cfm}. In fact, several searches for MMs were performed and reported together with searches for such electrically charged states. Nuclearites~\cite{Witten:1984rs,DeRujula:1984axn}, quirks~\cite{Kang:2008ea}, strangelets~\cite{Farhi:1984qu}, Q-balls~\cite{Coleman:1985ki,Kusenko:1997si}, black-hole remnants~\cite{Koch:2005ks,Hossenfelder:2005ku}, supersymmetric long-lived particles~\cite{Mavromatos:2016ykh,Sakurai:2019bac,Felea:2020cvf,Acharya:2020uwc}, D-particles~\cite{Ellis:2009vq,Shiu:2003ta,Mavromatos:2010nk,Elghozi:2015jka}, scalars in radiative-neutrino-mass models~\cite{R:2020odv,Hirsch:2021wge}, and many others fall into this category~\cite{MoEDAL:2014ttp}. 

MACRO performed a search for nuclearites, nuggets of strange quark matter surrounded by electrons, with the scintillator and NTD subdetectors. These objects may form part of cold dark matter~\cite{Bakari:2000wa}. No nuclearite candidate was found and a flux limit of $2.7 \times 10^{-16}~\cmssr$ was set for $\beta = 2 \times 10^{-3}$ for an isotropic flux of nuclearites~\cite{MACRO:1999ijn}. SLIM also searched for down-going nuclearites and charged Q-balls setting flux bounds at the level of  $1.3 \times 10^{-15}~\cmssr$~\cite{SLIM:2008bwg}.

The ANTARES neutrino telescope is sensitive to nuclearites  through
the black-body radiation emitted along their path. A search for heavy and slow moving $(\beta = 10^{-3})$ nuclearites with nine years of data led to flux upper limits for masses up to $10^{17}~\gev$ at the level of $\sim 5 \times 10^{-17}~\cmssr$~\cite{ANTARES:2022bqu}. These are the first upper limits on nuclearites established with a neutrino telescope and the most stringent ever set for Galactic velocities. These bounds can be improved by the KM3NeT~\cite{KM3Net:2016zxf}, especially in the high-mass region~\cite{Paun:2023pcz}.

Beside MMs, the ATLAS search described in Section~\ref{sc:atlas} has also set bounds on high-electric-charge objects (HECOs), excluding electric charges $20 \leq |z| \leq 100$, for masses up to 3.1~\tev~\cite{ATLAS:2023esy}, considering both DY and PF processes~\cite{Song:2021vpo}. MoEDAL, on the other hand, analysed a prototype array of Makrofol\textsuperscript{\textregistered} foils exposed to $\sqrt{s}=8~\tev$ $pp$ collisions, placing limits on HECO DY production~\cite{MoEDAL:2021mpi}. The analysis was repeated with the full Run-2 MoEDAL detector, comprising main NTDs and the HECC, exposed to higher integrated luminosity at $\sqrt{s}=13~\tev$ $pp$ collisions. Limits on HECOs produced via the DY and PF processes were placed for electric charges in the range $10e$ to $400e$ and masses up to $\sim3.9~\tev$~\cite{MoEDAL:2023ost}. A  plot summarising ATLAS and MoEDAL constraints on HECOs is available in Ref.~\cite{Mitsou:2026pdt}.

The high electric charge of HECOs, just like MMs, makes the photon--HECO and $Z$-boson--HECO coupling too large for perturbation theory to be valid for charges higher than $\sim10e$. Resummation techniques, discussed in Section~\ref{sc:resum}, offered a solution to this shortcoming for MMs~\cite{Alexandre:2019iub,Alexandre:2026auj}. Similar DS schemes have been applied to spin-0 and spin-\hf HECOs for both DY and PF processes.  The striking difference, though, is that in the HECO case, resummation leads to an \emph{enhancement} of the DY and PF production cross section, which in turn results in higher mass limits than the tree-level ones assumed by experimental collaborations in searches for such states~\cite{Alexandre:2023qjo,Alexandre:2024pbs,Musumeci:2025tuw}.

ATLAS and MoEDAL have considerable potential to discover generic electrically charged scalars and fermions in the range $1e$ to $6e$ in the High-Luminosity LHC (HL-LHC) runs~\cite{Altakach:2022hgn}. For MoEDAL, this is possible by analysing exposed CR-39\textsuperscript{\textregistered} foils, the calibration of which was recently completed~\cite{Kalliokoski:2025sog}. The analysis of this NTD type, which offers sensitivity to charges as low as $z/\beta=5$, is currently underway for stacks deployed in Run-2  $pp$ collisions.

It should be further noted that ATLAS cannot differentiate monopoles from HECOs, since the corresponding analysis is only sensitive to the ionisation signal.\footnote{An unambiguous MM-induced signal in ATLAS and CMS could come from an analysis based on the characteristic parabolic trajectory of a magnetically charged particle in the presence of a solenoid magnet, as discussed in Section~\ref{sc:coll-direct}. \label{fn:parabola}} In the MoEDAL NTDs, on the other hand, etch-pits from magnetic charges would be different than those from electric charges (as well as the their evolution in neighbouring foils), providing a clear signal of MMs, as discussed in Section~\ref{sc:ntd}. 

\section{Cosmic monopoles meet colliders}\label{sc:cosmic-colliders}

The MM-flux constraints discussed in Section~\ref{sc:cosmics} and presented in Figs.~\ref{fig:nova},~\ref{fig:antares} and~\ref{fig:auger-rice-hess} are expressed as a function of the MM velocity, usually treated as a free parameter due to the ambiguity in the computation of the acceleration before the MMs arrive at Earth. This picture is changing lately with developments with dedicated studies on the generation and acceleration of MMs before reaching Earth's surface.

Monopole production from collisions of cosmic rays bombarding the atmosphere has been considered as a MM source independent of cosmology, thus evading associated uncertainties. This approach supplies an irreducible MM flux for all terrestrial experiments. Using historic data from RICE~\cite{Hogan:2008sx}, AMANDA-II~\cite{Abbasi:2010zz}, and SLIM~\cite{Balestra:2008ps} experiments together with MM flux estimates from atmospheric cosmic ray collisions, bounds on the MM production cross section of in the \mbox{$\sim5$--100~\tev} mass range have been established and contrasted with collider constraints~\cite{Iguro:2021xsu}. 

\begin{figure}[htb]
\centering
\includegraphics[width=0.75\textwidth]
{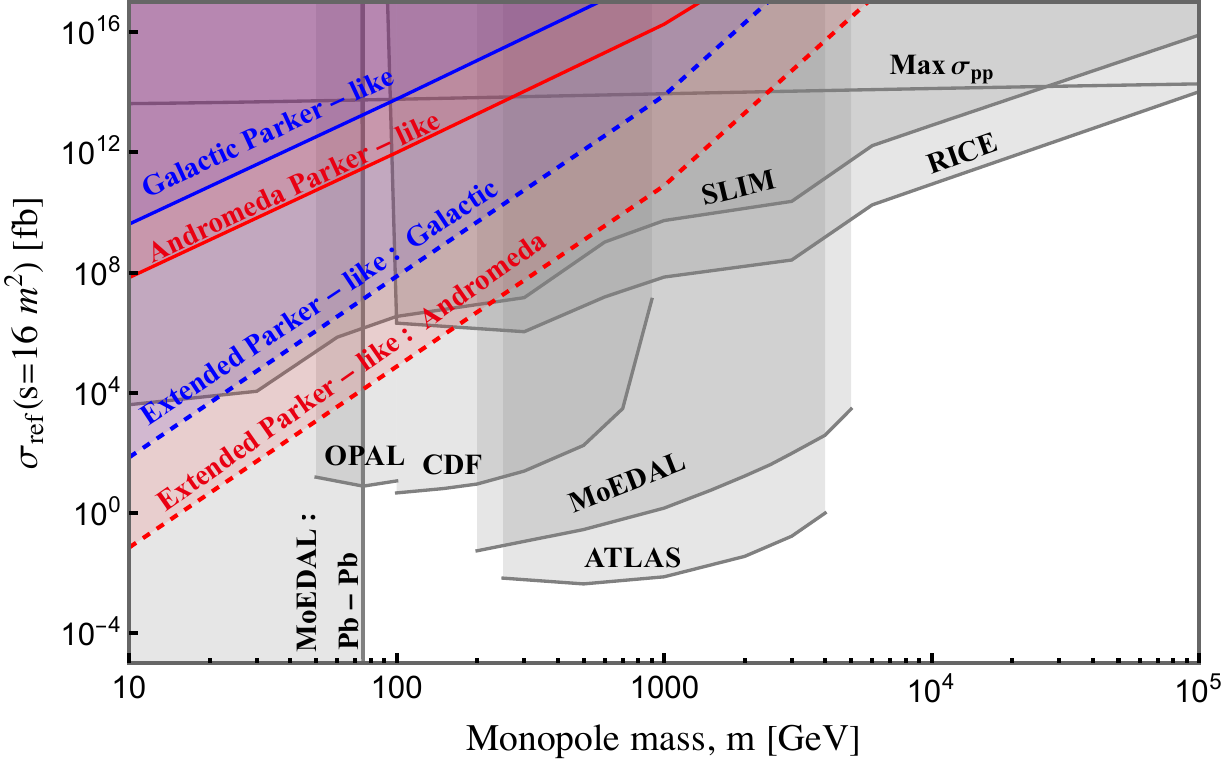}
\caption{
Summary of MM bounds on the reference cross-section $\sigma_\text{ref}$ from Parker-like and extended Parker-like constraints for a seed field of $B_{0} = 10^{-20}$~G~\cite{Iguro:2024oml}. Existing bounds (grey) are obtained from the combination of collider constraints (OPAL~\cite{OPAL:2007eyf}, CDF~\cite{CDF:2005cvf},  MoEDAL~\cite{MoEDAL:2019ort}, ATLAS~\cite{ATLAS:2019wkg}), as well as the RICE~\cite{Hogan:2008sx} and SLIM~\cite{Balestra:2008ps} data reinterpreted in Ref.~\cite{Iguro:2021xsu}. The very strong bounds on MM mass $\lesssim 75~\gev$ arise from the Schwinger-produced MMs in heavy-ion collisions~\cite{MoEDAL:2021vix,MoEDAL:2024wbc}. The maximum total cross section $\sigma_{pp \to X}$, as parameterised by the COMPETE collaboration~\cite{COMPETE:2002jcr} is also shown. From Ref.~\cite{Iguro:2024oml}.} 
\label{fig:cosmic-collider}
\end{figure}

Equally independent of primordial considerations would the  MM formation in interactions of cosmic rays and interstellar medium (ISM), leading to a broad spectrum of observational repercussions. Such MMs could significantly hinder the formation and sustainability of galactic magnetic fields, leading to generalised Parker bounds, e.g.\ from the Milky Way and Andromeda, as shown in Fig.~\ref{fig:cosmic-collider}. Unlike the Parker limits discussed in  Section~\ref{sc:parker}, these bounds are \emph{independent} of MM production in the early Universe. 
Moreover, in a similar fashion as with atmospheric MMs~\cite{Iguro:2021xsu}, $\beta$-dependent constraints on MM-flux have been recasted to bounds versus MM mass, making readily their comparison with collider limits obtained, e.g.\ from MoEDAL and ATLAS. These results are presented in Fig.~\ref{fig:cosmic-collider} in terms of upper limits on the reference cross-section $\sigma_\text{ref}$~\cite{Iguro:2024oml}.

Studies of MM acceleration in cosmic magnetic fields~\cite{Kobayashi:2021des,Kobayashi:2022qpl,Perri:2023ncd} offer a different avenue to connect collider searches for MMs with astrophysical conditions. A self-consistent estimate of MM acceleration in the magnetic fields in the Milky Way Galaxy and in the intergalactic space~\cite{Perri:2023ncd}, and their deceleration in Earth, combined with astrophysical constraints, such as the Parker bound (cf.\ Section~\ref{sc:parker}), allowed the recasting of existing MM-flux limits and projections in terms of the MM mass.  Specifically, this method was applied to terrestrial experiments, such as the Pierre Auger Observatory, IceCube, MACRO, and the upcoming CTAO, and was compared to MoEDAL results from LHC~\cite{Perri:2025qpg}. This result highlights the role of MMs as messengers of cosmic magnetic fields, and opens up the possibility of using MM experiments to probe intergalactic magnetic fields.

The aforementioned considerations allowed to present MM constraints from the Cosmos, discussed in Section~\ref{sc:direct} on equal footing with bounds from collider experiments, presented in Section~\ref{sc:colliders}. Such developments showcase the revived interest in MMs and their role in interconnections between a broad range of (astro)physical phenomena.

\section{A look to the future}\label{sc:future}

In the collider front, ATLAS pursues HIP searches using Run~3 data, however, the removal of the TRT during the upgrade for HL-LHC makes the future of such analyses questionable, requiring a change in the search strategy. Parabolic tracks\footref{fn:parabola} is a potential method for both ATLAS and CMS to explore. 
MoEDAL continues searching for MMs with detectors exposed to $pp$ and heavy-ion collisions using improved configurations of its detection systems~\cite{MoEDAL-MAPP:2022kyr}. 

ALICE~\cite{ALICE:2008ngc} has entered this endeavour with a proposal to look for MMs, and HIPs in general, in its time projection chamber. The method relies on the negatively polarised common-mode signal as a hardware trigger in the readout electronics, enabling the recording of digitised raw data without loss of information, which is crucial for the precise energy loss and mass measurements. A study on large energy depositions, typical of a MM, was promising~\cite{Arslandok:2024hhw}.

The option of analysing decommissioned parts of LHC experiments using the induction technique to search for trapped MMs has already delivered its first outcomes in a scanning of the CMS beam pipe by the MoEDAL Collaboration~\cite{MoEDAL:2024wbc}. In this context, other decommissioned parts from past LHC runs, in particular in view of the planned upgrade of experiments for HL-LHC, may serve as potential MM trappers to be cut and scanned via a SQUID in search for MMs. 

The Schwinger mechanism offers good prospects for detecting monopoles at future hadron colliders such as FCC-hh~\cite{FCC:2025lpp} and SPPC~\cite{CEPCStudyGroup:2023quu}, expected to reach $pp$ collision energy of 100~\tev and 125~\tev, respectively. The reach of such machines in detecting MMs in Pb--Pb collisions~\cite{dEnterria:2022sut} has been studied assuming an ATLAS-like, general-purpose experiment capable of a 50\% detection efficiency, and a HIP-optimised MoEDAL-like detector, both under a zero-background hypothesis. Both concepts will provide sensitivity to \tev-scale masses for MMs of charges 1\gd or higher~\cite{Gould:2024zed}.

Furthermore, MMs can be probed in colliders \emph{indirectly} in events with multiple photons in the final state. Promising prospects exist for searches of virtual-MM-induced diphoton production enhancements in future electron--positron colliders~\cite{Ellis:2022uxv}, photon--photon~\cite{Jikia:1993tc,Ginzburg:1999ej,Ginzburg:2020koq}, and muon colliders~\cite{Yang:2022ilt}. Similar experimental signatures may constrain monopolia~\cite{Epele:2012jn,Barrie:2016wxf}. In addition, MMs may bind deeply with neutral states, forming the so-called \emph{hideons}, characterised by a large magnetic charge and a magnetic moment. An interesting low-mass case, the \emph{magnetron}, an analog of an electron but with magnetic instead of electric charge, could potentially be explored at collider experiments~\cite{Fanchiotti:2024kxn}. 

In the cosmic front, neutrino telescopes, such as IceCube and KM3NeT~\cite{KM3Net:2016zxf} currently, and IceCube Upgrade~\cite{IceCubeCollaborationP:2025rpl} and IceCube-Gen2~\cite{3128622} in the future, will pursue the search for very fast MMs. Besides them, neutrino-oscillation experiments, such as \nova, currently in operation at Fermilab, and the Iron CALorimeter (ICAL) detector, to be built at the proposed India-based Neutrino Observatory (INO) facility~\cite{Dash:2014fba}, will be sensitive to MMs in the sub-relativistic range. Future neutrino experiments DUNE and Hyper-Kamiokande, can probe MMs via high-energy antiproton production and proton decay catalysis~\cite{Candela:2025gwp}.

A cosmic MM search experiment (Search for Cosmic Exotic Particles, SCEP) has been proposed, utilising a hybrid approach that combines radio-frequency atomic magnetometers and plastic scintillators. Such setup would allow for the collection of both the induction and scintillation signals generated by the passage of a MM, which provides acceptance to MM with velocity $\beta > 10^{-6}$ and masses larger than  $\sim 10^7~\gev$~\cite{SCEP:2024cir}.

Moreover, primordial MM may loose their kinetic energy in the atmosphere and drift toward the magnetic poles to eventually be trapped in ice. By scanning ice samples with a SQUID magnetometer, 1-\tev MMs of $\gamma \lesssim 6$ could be probed~\cite{Gould:2024zed}. The same scanning framework may be used on samples of industrially extracted ore, thus reviving earlier searches for MMs trapped in Earth-based rocks and deep-sea sediments.

MANDATE (Monopole And Nuclear Detector at Tibet Elevation) is a proposed large-area array of NTDs to be deployed at high altitude on the Tibetan Plateau. Building on the previous generation of dedicated searches, such as SLIM, MANDATE introduces the novel approach of operating in conjunction with a non-dedicated cosmic ray observatory, such as LHAASO, to maximise the sensitivity and scientific reach~\cite{Gould:2024zed}. The detector will search for cosmic MMs from the \tev to the GUT scale for fluxes well below the Parker bound. A pilot deployment is currently being commissioned on the territory of LHAASO to validate the full chain of detector preparation, deployment, data accumulation, and processing. A conceptual design report for the full-scale experiment is planned for publication in early 2027.

\section{Epilogue}\label{sc:summary}

The existence of magnetic monopoles, if confirmed experimentally,  would revolutionise our understanding of Electromagnetism, rendering the Maxwell equations formally fully symmetric. The Dirac electric-charge quantisation condition is an elegant consequence of the existence of MMs. Hence, it represents an extremely appealing motivation for studying and looking for isolated magnetic charges. The interest in MM searches at colliders and in the Cosmos has been revived in the past few years due to new theoretical proposals, phenomenological developments and experimental opportunities.

A wide range of theoretical microscopic models of MMs, ranging from string and GUT theories to (extensions of) the Standard Model, have been proposed. Theoretical scenarios predicting relatively light MMs, with masses of order of the electroweak scale, made composite MMs accessible, in principle, at colliders. Advances in semi-classical calculations \`a la Schwinger have led to MM production constraints in strong magnetic fields. Dyson--Schwinger-like resummation schemes have confronted the non-perturbativity of the large MM--photon coupling. 

Numerous MM searches have beed carried out for decades by utilising diverse detection techniques in both observational facilities and experiments in colliding beams. The CERN LHC, being the most powerful collider to-date, leads this effort via two complementary detection approaches applied by the ATLAS and MoEDAL experiments. ATLAS is has better sensitivity in low magnetic charges, while MoEDAL has set the stringent bounds in high charges. ALICE joins the effort with a proposal to use its time-of-flight chamber as passive detector. Multiphoton final states further constrain MMs indirectly through virtual MMs in loops and the decay of bound states.

Cosmologically created MMs are being constrained considering measuring physical properties of astrophysical objects and the survival of magnetic fields. Neutrino and cosmic-ray telescopes lead the search for cosmic MMs in the relativistic regime, whereas neutrino-oscillation-oriented and other dedicated experiments explore slower-moving MMs.  

Magnetic monopoles continue to fascinate scientists in the fields of Particle Physics and Cosmology. Exciting developments are being made in theoretical scenarios, while experimental advancements allow their exploration using diverse detection approaches. Future prospects keep looking promising for these elusive particles.

\begin{acknowledgement}
The author acknowledges support by the Spanish MICIN / AEI / 0.13039/50110001103 and the European Union / FEDER via the grants PID2021-122134NB-C21 and PID2024-158190NB-C21, and by the MICIU/AEI grant Severo Ochoa CEX2023-001292-S.
\end{acknowledgement}
%


\bibliographystyle{JHEP-vaso}
\bibliography{mitsou-monopoles.bib}

\end{document}